\documentclass[useAMS,usenatbib]{mn2e}
\usepackage{graphicx}
\usepackage{times}

\newcommand{\CIV}{\hbox{{\rm C}{\sc \,iv}}}

\newcommand{\FeII}{\hbox{{\rm Fe}{\sc \,ii}}}

\newcommand{\OVI}{\hbox{{\rm O}{\sc \,vi}}}
\newcommand{\NaI}{\hbox{{\rm Na}{\sc \,i}}}
\newcommand{\MgI}{\hbox{{\rm Mg}{\sc \,i}}}
\newcommand{\MgII}{\hbox{{\rm Mg}{\sc \,ii}}}
\newcommand{\CaII}{\hbox{{\rm Ca}{\sc \,ii}}}

\newcommand{\HI}{\hbox{{\rm H}{\sc \,i}}}

\newcommand{\lya}{\hbox{{\rm Ly}$\alpha$}}

\newcommand{\Ha}{\hbox{{\rm H}$\alpha$}}

\newcommand{\msun}{\hbox{M$_{\odot}$}}
\newcommand{\cmsq}{\hbox{cm$^{-2}$}}
\newcommand{\NHI}{\hbox{$N_{\rm HI}$}}
\newcommand{\ma}{\hbox{$\lambda 2796$}}
\newcommand{\mb}{\hbox{$\lambda 2803$}}
\newcommand{\mc}{\hbox{$\lambda 2852$}}
\newcommand{\fa}{\hbox{$\lambda 2600$}}
\newcommand{\LL}{\hbox{$\lambda\lambda$}}
\newcommand{\zem}{\hbox{$z_{\rm em}$}}
\newcommand{\zabs}{\hbox{$z_{\rm abs}$}}
\newcommand{\kms}{\hbox{${\rm km\,s}^{-1}$}}
\newcommand{\nn}{\nonumber}
\newcommand{\hkpc}{\hbox{$h^{-1}$~kpc}}
\newcommand{\hMpc}{\hbox{$h^{-1}$~Mpc}}
\newcommand{\dNdzdW}{\hbox{$\mathrm d^2 N/\mathrm d z/\mathrm d W_{\rm r}$}}
\newcommand{\dNdz}{\hbox{${\mathrm d N}/{\mathrm d z}$}}
\newcommand{\dNdl}{\hbox{${\mathrm d N}/{\mathrm d l}$}}

\newcommand{\Hgamma}{\hbox{$\frac{\Gamma(\frac{1}{2})\Gamma(\frac{\gamma-1}{2})}{\Gamma(\frac{\gamma}{2})}$}}
\newcommand{\EW}{\hbox{$W_{\rm r}^{\lambda2796}$}}
\newcommand{\Mh}{\hbox{$M_{\rm h}$}}

\newcommand{\nabstot}{\hbox{1806}}

\newcommand{\ngalsim}{\hbox{$\sim$~250,000}}
\newcommand{\ngalused}{\hbox{242,620}}

\newcommand{\Aauto}{\hbox{$0.186\pm 0.012$}}
\newcommand{\Bauto}{\hbox{$-1.023\pm 0.035$}}
\newcommand{\Across}{\hbox{$0.143\pm 0.012$}}
\newcommand{\Bcross}{\hbox{$-0.893\pm 0.043$}}

\newcommand{\Arel}{\hbox{$0.808 \pm 0.096$}}
\newcommand{\Arelcorr}{\hbox{$0.65\pm 0.08$}}
\newcommand{\Arelcorrwithsys}{\hbox{$0.65\pm 0.09$}}
\newcommand{\Arelcorrsys}{\hbox{$0.65 \pm 0.08 (\hbox{stat}) \pm {0.05} (\hbox{sys})$}}

\newcommand{\LRGmassrange}{\hbox{$11.94\pm {0.31}$}}  
\newcommand{\LRGmassrangesysnolrg}{\hbox{$11.94\pm {0.31} (\hbox{stat})  ^{+0.20}_{-0.19}$ (\hbox{sys})}}  
\newcommand{\LRGmassrangeallsys}{\hbox{$11.94\pm {0.31} (\hbox{stat})  ^{+0.24}_{-0.25}$ (\hbox{sys})}}  
\newcommand{\LRGmassrangesys}{\hbox{$11.94^{+0.39}_{-0.40}$}}  
\newcommand{\LRGmassrangehigh}{$12.5_{-0.3}^{+0.3}$}
\newcommand{\LRGmassrangelow}{$ 11.3^{+0.4}_{-0.4}$}
\newcommand{\overestimate}{\hbox{$25\pm10$}}
\newcommand{\massratio}{\hbox{$\sim 20$--40}}

\def\bsp_small{\vspace{0.5cm}\small\noindent This paper
has been typeset from a \TeX / \LaTeX\ file prepared by the author.}

\title[Strong $\bmath{z \!\simeq\! 0.5}$ {\MgII} absorbers]
{New perspectives on strong $\bmath{z \!\simeq\! 0.5}$ {\MgII} absorbers: 
are halo-mass and equivalent width anti-correlated?
}
  \author[N.~Bouch\'e, M.~T.~Murphy, C.~P\'eroux, I.~Csabai, V.~Wild]{Nicolas
  Bouch\'e$^1$\thanks{E-mail: nbouche@mpe.mpg.de (NB)}, Michael T.~Murphy$^2$,
  C\'eline P\'eroux$^3$,  Istv\'an Csabai$^4$, Vivienne Wild$^5$\\
   $^1$Max Planck Institut f\"ur extraterrestrische Physik,
Giessenbachstrasse, Garching D-85748, Germany\\
  $^2$Institute of Astronomy, University of Cambridge, Madingley Road,   Cambridge CB3 0HA, UK\\
  $^3$European Southern   Observatory, Karl-Schwarzschild-Str 2, D-85748 Garching, Germany\\
  $^4$Dept. of Physics, E\"otv\"os Lor\'and University, Budapest, Pf. 32, H-1518, Budapest, Hungary\\
  $^5$Max Planck Institut f\"ur Astrophysik, Karl-Schwarzschild-str 1, D-85748 Garching, Germany
}

\begin{document}

\date{Accepted 2006 June 13.
      Received 2006 June 12;
      in original form 2006 April 15}

\pagerange{\pageref{firstpage}--\pageref{lastpage}}
\pubyear{2006}

\maketitle

\label{firstpage}

\begin{abstract}
  We measure the mean halo-mass of $z\!\simeq\!0.5$ \MgII\ absorbers
  using the cross-correlation (over co-moving scales
  0.05--13$h^{-1}{\rm \,Mpc}$) between \nabstot\ \MgII\ quasar
  absorption systems and \ngalsim\ Luminous Red Galaxies (LRGs), both
  selected from the Sloan Digital Sky Survey Data Release 3.  The
  \MgII\ systems have \ma\ rest-frame equivalent widths $\EW\ga
  0.3$\,\AA.  From the ratio of the \MgII--LRG cross-correlation to
  the LRG--LRG auto-correlation, we find that the bias ratio between
  \MgII\ absorbers and LRGs is $\overline b_{\rm \MgII}/\overline
  b_{\rm LRG}=\Arelcorr$, which implies that the absorber
  host-galaxies have a mean halo-mass $\massratio$ times smaller than
  that of the LRGs; the \MgII\ absorbers have haloes of mean mass
  $\langle \log \Mh (\msun)\rangle=\;$\LRGmassrangeallsys. We
  demonstrate that this statistical technique, which does not require
  any spectroscopic follow-up, does not suffer from contaminants such
  as stars or foreground and background galaxies.  Finally, we find
  that the absorber halo-mass is anti-correlated with the equivalent
  width. If \MgII\ absorbers were virialized in galaxy haloes a
  positive \Mh--\EW\ correlation would have been observed since \EW\
  is a direct measure of the velocity spread of the \MgII\ sub-components.
  Thus, our results demonstrate that the individual clouds of a \MgII\ system are {\it not}
  virialized in the gaseous haloes of the host-galaxies. We review past
  results in the literature on the statistics of \MgII\ absorbers and
  find that they too require an \Mh--\EW\ anti-correlation.  When
  combined with measurements of the equivalent width distribution
  (\dNdzdW), the \Mh--\EW\ anti-correlation naturally explains why
  absorbers with $\EW\ga2$\,\AA\ are not seen at large impact
  parameters.  We interpret the \Mh--\EW\ anti-correlation within the
  starburst scenario where strong \MgII\ absorbers are produced by
  supernovae-driven winds.
\end{abstract}

\begin{keywords}
cosmology: observations --- galaxies: evolution ---  galaxies: haloes ---
 quasars: absorption lines
\end{keywords}

\section{Introduction}
The connection between quasar (QSO) absorption line (QAL) systems and
galaxies is crucial to our understanding of galaxy evolution since
QALs provide detailed information about the physical conditions of
galaxy haloes out to large impact parameters ($\rho\!>\!100{\rm \,kpc}$)
with no direct dependence on the host galaxy luminosity.  \MgII\
absorbers are ideal for this purpose as the \MgII\LL\ 2796,2803
doublet can be detected from $z\simeq0.3$ to $z\simeq 2.2$ at optical
wavelengths. Since the ionization potential of \MgI\ is less than
13.6\,eV but the ionization potential of \MgII\ is greater than
13.6\,eV, \MgII\ absorbers trace cold gas. In fact \MgII\ absorbers
with equivalent widths $\EW\ga0.03$\,\AA\ have been shown to be
associated with \HI\ absorbers covering five decades in \HI\ column
density ($N_{\HI}$) \citep[e.g.][]{PetitjeanP_90a}, including
sub-Lyman limit systems \citep{ChurchillC_99a,ChurchillC_00a}, Lyman
limit systems \citep[e.g.][]{BergeronJ_86a,SteidelC_92a} and damped
\lya\ systems \citep[DLAs;
e.g.,][]{LeBrunV_97a,RaoS_00a,BoisseP_98a,ChurchillC_00b,RaoS_06a}, which means
that a large range of galactic environments are likely to be sampled.
For example,  strong absorbers with $\EW\!\geq\!0.3$\,\AA\ are known to be
generally associated with galaxies
\citep{LanzettaK_90a,BergeronJ_91a,BergeronJ_92a,SteidelC_92a,DrinkwaterM_93a,SteidelC_94a}.
These groups have shown that
galaxies responsible for the \MgII\ host-galaxies have luminosities
consistent with normal field $0.7L^*_B$ galaxies, and
\citeauthor{SteidelC_94a} showed that, from their average colour, they
are on average late-type (Sb) galaxies, a result reproduced by
\citet{ZibettiS_05a}.

Medium-resolution spectroscopy of the \MgII\ absorbers quickly
revealed that such absorbers are composed of several sub-components
\citep{BergeronJ_86a,TytlerD_87a,PetitjeanP_90a} and
that the number of sub-components    strongly
correlates with equivalent width \citep{PetitjeanP_90a}.
Building on this
work, \citet{ChurchillC_97a} showed that the mean Doppler width of
individual components is 5\,\kms, with an rms of comparable magnitude,
using high resolution spectra (${\rm FWHM} \simeq 6$\,\kms).  This
corresponds to a thermal temperature of $\sim$30,000~K.
\citet{ChurchillC_97a} also directly constrained the turbulent
component to be $\la 2$\,\kms.  \citet{ChurchillC_03a}, using
high-resolution spectra, confirmed that equivalent width does strongly
correlate with the number of sub-components as shown by
\citet{PetitjeanP_90a}. This correlation arises because a large
equivalent width can only be produced by more components spread over a
large velocity range since extremely few components with large Doppler
widths are seen. The equivalent width, \EW, should therefore be correlated
with the velocity range, $\Delta v$, covered by the sub-components
and, indeed, this is observed to be the case \citep[e.g.][see their
figure 3]{EllisonS_06a}. The velocity range for strong \MgII\ systems
 are very large, from 50 to 400~\kms.
If the individual
clouds were virialized within the haloes of the host galaxies, $\Delta v$
would represent the velocity dispersion of the many clouds in the host-galaxy
halo. In this case, $\Delta v$ would be directly related to the mass of the host-galaxy.
Since a larger \EW\ is achieved by having a
larger number of clouds spread over a larger $\Delta v$, 
$\EW$ should also be positively correlated with the mass of the host galaxies {\it if}
 the individual
clouds are virialized within the haloes of the host galaxies.

Among the few correlations observed between the host-galaxy properties
and the absorber properties, \citet{LanzettaK_90a} noted a significant
anti-correlation between \MgII\ equivalent width and the impact
parameter distribution from the sample of \citet{BergeronJ_91a}.
While \citet{BergeronJ_91a} argued that the anti-correlation was not
significant, it was clearly seen in the sample of \citet{SteidelC_92a}
and \citet{SteidelC_95b}: absorbers with $\EW \ga 2$\,\AA\ are observed
at small impact parameters but do not exist at large impact parameters
($\rho\ga 50$\,kpc).
This is puzzling as large \EW\ absorbers at large impact parameters
ought to be easily detectable.

Only a slight correlation is observed between absorption cross-section
and galaxy luminosity. \citet{SteidelC_93a} and \citet{SteidelC_95b}
showed that the cross-section of \MgII\ absorbers with
$\EW\!\geq\!0.3$\,\AA\ has radius $R_{\rm
  phys}\!\sim\!40\;h^{-1}{\rm \,kpc}$ (physical, $70\;h^{-1}{\rm \,kpc}$
co-moving) from the observed absorber--host-galaxy impact parameter
distribution.  Furthermore, they concluded that the cross-section
radius only slightly increases with luminosity, $R_{\rm
  phys}\propto (L_K/L_K^*)^{0.2}$.

The only other known correlation between galaxy properties and
absorber properties is that recently discovered between \EW\ and the
galaxy morphological asymmetry \citep{KacprzakG_05a,KacprzakG_06a}.
\citeauthor{KacprzakG_06a} searched for correlations between gas
properties, host galaxy impact parameters, inclinations (\&\ position
angles) and morphological parameters.  Among those parameters, they
only found a 3.2-$\sigma$ correlation between \EW\ and the asymmetry
in the host galaxy morphology as measured from the residuals of 2D
light profile fits.

From studies of individual absorbers, some recent work that compared
the absorption kinematics with the host galaxy rotation curve favour
the idea that strong \MgII\ clouds arise in galactic outflows
\citep{BondN_01a,EllisonS_03a}.  On the other hand,
\citet{SteidelC_02a} showed that, in 4 out of 5 \MgII\ absorbers
(selected to be edge-on and aligned towards to QSO), the absorption
kinematics are consistent with rotation being dominant for the
absorbing-gas kinematics. However, a simple extension of the 
rotation curve fails.

Many authors have tried to put QALs in the context of theoretical
models since \citet{BahcallJ_69a} who proposed that the metal
absorption lines are produced by gaseous haloes of intervening galaxies
with a large cross-section (up to 100~kpc).  Later, \citet{YorkD_86a}
argued that QALs arise mainly in the haloes of gas rich Magellanic-type
dwarfs.  \citet{MoH_96a} followed the ideas of \citet{BahcallJ_69a}
and produced a detailed model in which \MgII\ absorbers are signatures
of in-falling (photo-ionized) cold gas embedded in a $T\simeq10^6$~K
gas halo. In this model, the in-falling cold gas should be virialized.
\citet{MallerA_04a} reached similar conclusions.

Despite these numerous past results, a fundamental question remains:
What is the physical nature of strong \MgII\ absorbers?  Here we
constrain one important physical property of \MgII\ absorbers --
namely the halo-mass of the host galaxies -- {\it statistically} by
studying the clustering of galaxies around the absorbers.  Clustering
studies of metal-line systems, such as the absorber--absorber
auto-correlation in velocity space, have been used for some time
\citep[e.g.][]{SargentW_88a,SteidelC_92a,ChurchillC_97a,CharltonJ_98a}.
This technique measures the line-of-sight clustering and therefore
suffers from strong peculiar velocity effects.  Numerical models are
then required to infer the physical properties (halo mass, halo sizes,
etc.) of the host galaxies \citep[see][for a recent
example]{ScannapiecoE_06a}.

In order to avoid these limitations, we choose to cross-correlate
\MgII\ absorbers with `field' galaxies tracing the two-dimensional
large scale structure as in \citet{BoucheN_04c}, \citet*{BoucheN_04a},
\cite{BoucheN_05b}, \citet{CookeJ_06a} and \citet{Ryan-WeberE_06a}.
Specifically, we measure the amplitude {\it ratio} of the
absorber--galaxy cross-correlation to the galaxy--galaxy
auto-correlation. In hierarchical galaxy formation scenarios this is a
direct measure of the ratio between the bias of the absorber
host-galaxies and of the field galaxies used. From this bias ratio,
the mass of the absorber host-galaxies can be inferred.

In this paper, we extend our DR1 results \citep{BoucheN_04a} by
cross-correlating $\nabstot$ \MgII\ absorbers with $\ngalsim$ Luminous
Red Galaxies (LRGs), both selected from the Sloan Digital Sky Survey
(SDSS) Data Release 3 \citep[DR3;][]{AbazajianK_05a}, over
$\sim\!800$\,square degrees. Thanks to the large SDSS survey, this is
a leap forward from other clustering studies such as
\citet{WilligerG_02a} and \citet{HainesC_04a} which cover a few square
degrees.

This paper is organized as follows. In Section~\ref{section:sample},
we summarise the selection of our \MgII\ absorbers and LRGs.  In
Section~\ref{section:method}, we describe our method to measure the
halo-mass.  The results are presented in
Section~\ref{section:results}.  We test these results against numerous
past results on \MgII\ absorbers in Section~\ref{section:discussion}
and discuss a physical interpretation of our results in
Section~\ref{section:interpretation}. Our main results are summarised
in Section~\ref{section:summary}.  For those familiar with
absorber--galaxy clustering analyses, a quick read of this paper
comprises Fig.~\ref{fig:xcorr:050505:2halo} and
Fig.~\ref{fig:xcorr:EW}, followed by
Section~\ref{section:interpretation}.  A critical reader should focus
on the several consistency tests we performed
(Figs.~\ref{fig:xcorr:test:z_off} \& \ref{fig:xcorr:test:Wz}) and on
the discussion in Section~\ref{section:discussion}.

We adopt $\Omega_{\rm M}\!=\!0.3$, $\Omega_\Lambda\!=\!0.7$ and
$H_0\!=\!100 h\,\kms{\rm \,Mpc}^{-1}$ throughout. Thus, at
$z\!=\!0.6$, $1\arcsec$ corresponds to $7.44h^{-1}{\rm \,kpc}$ and
$\delta z\!=\!0.1$ corresponds to $216 h^{-1}{\rm \,Mpc}$, both in
co-moving coordinates.

\section{Sample definitions}
\label{section:sample}

\subsection{\MgII\ absorbers}

\begin{figure*}
  \centerline{\includegraphics[width=130mm]{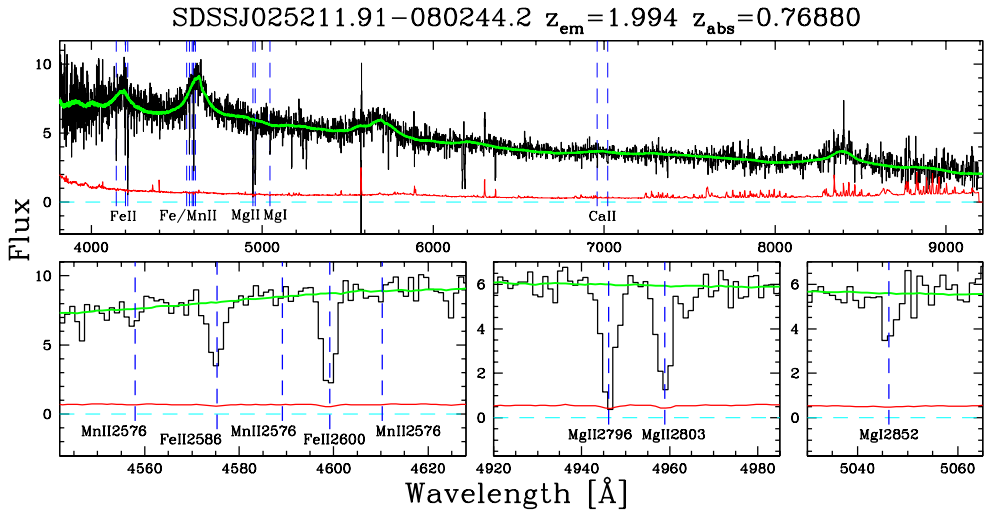}}
  \caption{SDSS spectrum of QSO SDSSJ025211.91$-$080244.2 containing a
    strong \MgII\ absorber at $z_{\rm abs}=0.7688$. The upper panel
    shows the entire spectrum (black histogram) with our PCA
    reconstruction of the QSO continuum (solid line).  The error array
    is shown by the continuous line just above the zero flux level.
    The vertical dashed lines mark several transitions at the
    absorption redshift, including the \FeII\ $\lambda2600$ and the
    \MgI\ $\lambda2852$ lines that we used to confirm the \MgII\
    doublet. These transitions are detailed in the lower panels.
    Detailed plots for all of our 1806 \MgII\ absorbers can be viewed
    at {http://www.ast.cam.ac.uk/$\sim$mim/pub.html}.  }
\label{fig:example}
\end{figure*} 

The algorithm to select \MgII\ absorbers from SDSS/DR3 differs in
several ways from the algorithm used for SDSS/DR1 in
\citet{BoucheN_04a}, the most important of which is the method for
estimating the QSO continuum. In \citet{BoucheN_04a} we used a series
of overlapping polynomial fits to small Sections of the continuum.
While this provides a reliable continuum in most cases, it does not
perform well near sharp QSO emission lines, particularly when
absorption lines -- possibly the target \MgII\ lines -- are imprinted
over the emission. To alleviate this problem in the current SDSS/DR3
analysis, we used principal component analysis (PCA) reconstructions
of the QSO continua. Full details of the method are reported in
\citet[][see their appendix]{WildV_06a}.

The features of the PCA algorithm most important for the present paper
follow. Eigenspectra were generated in 4 QSO emission redshift bins to
reduce the amount of data `missing' due to the differing wavelength
coverage of each spectrum: 0.005--0.458, 0.381--0.923, 0.822--1.537,
1.410--2.179, 2.172--3.193.  An iterative procedure was used to
identify and remove quasars showing broad absorption lines (BALs)
during the creation of the PCA eigenspectra.  This improves the
continuum reconstruction of non-BAL quasar spectra, by removing from
the input sample features which vary greatly in a small number of
objects.  Fig.~\ref{fig:example} shows an example SDSS spectrum with
its PCA continuum.


Having established a continuum, candidate \MgII\ absorbers are
identified as follows. We searched for intervening \MgII\ absorbers
from $\zabs= 0.37$ to $\zabs = 0.8$ (see Section~\ref{section:LRGs}).
The low redshift cut arises from the fact that SDSS QSO spectra begin
at $\approx$3800\AA, and the high redshift cut was imposed since this
is where LRGs drop below the SDSS magnitude limits. In addition, we
require all \MgII\ systems to be above the \lya\ QSO emission line.
QSOs above $\zem = 3.193$ are therefore not considered.  All pixels
above 1250\,\AA\ in the QSO rest-frame and $>3000\,\kms$ blue-wards of
the MgII emission line are tested for \MgII\ \ma\ absorption. At each
pixel, putative \MgII\ \ma\ lines are characterized using a method
similar to that detailed by \citet{SchneiderD_93a}. For initial line
detection, the rest-frame \MgII\ \ma\ equivalent width and 1-$\sigma$
detection limit are defined using the spectrograph instrumental
profile (IP) as a weighting function and using pixels in a 7\,\AA\
window in the putative absorber's rest-frame. The IP is assumed to be
a Gaussian of width ${\rm FWHM} = 160\,\kms$. If this estimate of the
\MgII\ \ma\ equivalent width is $\ge 0.3$\,\AA\ and is significant at
$\ge 8\,\sigma$ then the putative absorption redshift is estimated
from a parabolic interpolation of the equivalent widths of the current
pixel and its 2 neighbouring pixels. The equivalent width of the
\MgII\ \mb\ line is estimated in a similar way based on the \ma\
redshift. If the \mb\ equivalent width is significant at $\ge
3\,\sigma$ then the system is flagged as a candidate \MgII\ absorber.
Spurious candidates are removed by visually inspecting each \MgII\
candidate. The most common mis-identification is broad \CIV\
absorption near the \CIV\ emission line. Adopting a conservative
approach, we rejected any candidates which did not show absorption in
either \FeII\ \fa\ or \MgI\ \mc\ at $\ge 1\,\sigma$ significance.

For each candidate we derive a refined estimate of the \MgII\ \ma\
equivalent width using a Gaussian fit to the absorption as a weighting
function (c.f.~the IP-weighting above). This new, somewhat more
optimal, estimate is referred to throughout this paper as the measured
\MgII\ \ma\ equivalent width, \EW, for the system. Equivalent widths,
with a similar Gaussian-fit weighting, are also derived for a variety
of other commonly observed transitions (see Table \ref{table:mg}).

With the above algorithm we detected and visually confirmed \nabstot\
\MgII\ absorbers in SDSS/DR3. Figure~\ref{fig:zspec} (left panel) shows the distribution
of the equivalent width of our \MgII\ absorbers.
Table \ref{table:mg} is an excerpt from
the catalogue of absorbers which is available in its entirety in the
electronic edition of this paper and from an on-line catalogue at
http://www.ast.cam.ac.uk/$\sim$mim/pub.html. Figure \ref{fig:example}
shows an example \MgII\ absorption system. Similar plots are available
for all absorption systems in the on-line catalogue.

\begin{table}
\begin{center}
\vspace{-1mm}

\caption{Catalogue of \nabstot\ \MgII\ absorbers from the SDSS DR3
  with $0.37\!\leq\!z_{\rm abs}\!\leq\!0.80$. The J2000 name, QSO and
  absorption redshifts and the measured $W_{\rm r}$ for \MgII \LL 2796
  \& 2803, \MgI\ \mc\ and \FeII\ \fa\ are given. Here we show only a
  small sample from the full table which is available in the
  electronic edition of this paper and from
  http://www.ast.cam.ac.uk/$\sim$mim/pub.html. Full name designations,
  and statistical uncertainties in $W_{\rm r}$ are given in the
  electronic version, together with $W_{\rm r}$ measurements for the
  following transitions: 
  Zn{\sc \,ii} \LL2026/2062, 
  Cr{\sc \,ii} $\lambda$2056/$\lambda$2062/$\lambda$2066, 
  Fe{\sc \,ii} $\lambda$2344/$\lambda$2374/$\lambda$2382/$\lambda$2586, 
  Mn{\sc \,ii} $\lambda$2576/$\lambda$2594/$\lambda$2606, 
  Ti{\sc \,ii} $\lambda$3242/$\lambda$3384, 
  Ca{\sc \,ii} \LL3934/3969 and 
  Na{\sc \,i}  \LL5891/5897.
 }
\vspace{-4mm}
\label{table:mg}
\begin{tabular}{lcccccc}\hline
 & & &\multicolumn{4}{c}{Rest equivalent width $\left [ {\rm \AA} \right ]$ }\\
\multicolumn{1}{c}{SDSSJ}&\multicolumn{1}{c}{\zem}&\multicolumn{1}{c}{\zabs}&
\multicolumn{1}{c}{2796}&\multicolumn{1}{c}{2803}&\multicolumn{1}{c}{2852}&\multicolumn{1}{c}{2600}\\\hline
160530$+$393116 & 1.083 & 0.4969 & 1.04 & 0.79 & 0.04 & 0.19\\
160726$+$471251 & 1.816 & 0.4974 & 1.30 & 1.41 & 0.03 & 0.59\\
171726$+$654542 & 1.215 & 0.4974 & 3.29 & 3.39 & 0.22 & 1.39\\
112719$+$654143 & 1.250 & 0.4977 & 0.53 & 0.32 & 0.17 & 0.24\\
\hline
\end{tabular}
\end{center}
\end{table}

\subsection{Luminous red galaxies}\label{section:LRGs}

\begin{figure*}
\centerline{\includegraphics[width=70mm]{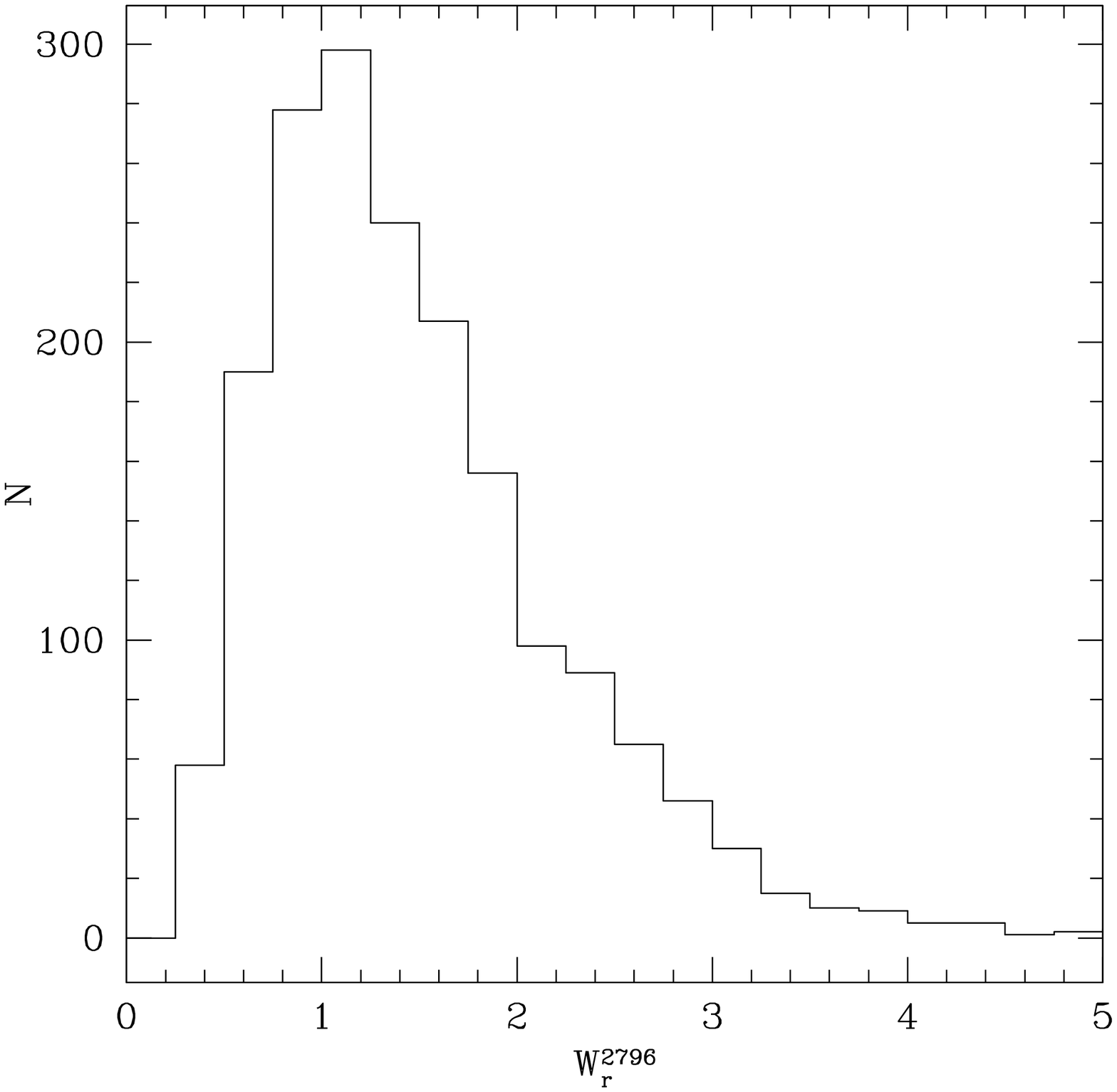}
\includegraphics[width=70mm]{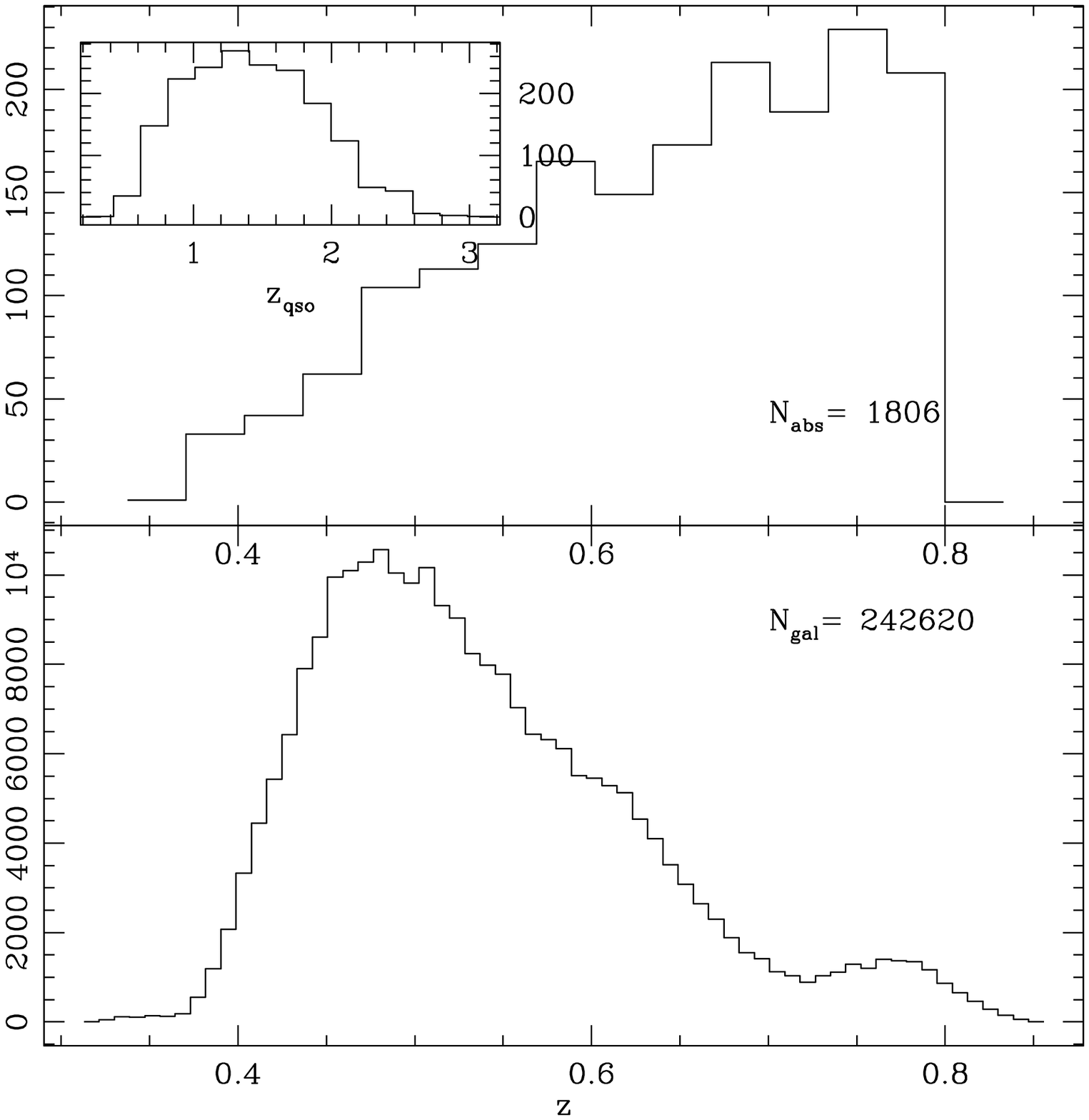}}
\caption{{\it Left}: Rest-frame equivalent width (\EW) distribution
of our 1806 \MgII\ systems.  {\it Right}: Redshift distribution of the
LRGs (bottom) and \MgII\ systems (top). The inset shows the
distribution of QSO emission redshifts.}
\label{fig:zspec}
\end{figure*}

\citet{EisensteinD_01a} \citep[see also][]{ScrantonR_04a} presented
colour criteria specifically designed to select luminous massive early
types both locally and at $z>0.4$, i.e.~much beyond the volume of the
SDSS Main sample. Because of their association with luminous
($M_r\sim-21.5$), massive haloes and spectral uniformity, LRGs are
excellent probes of the large scale structure as proven by the
detection of the baryon oscillations by \citet{EisensteinD_05a}.

For each \MgII\ absorber, galaxies meeting the following criteria
\citep[following][]{ScrantonR_04a} were extracted from the SDSS DR3
galaxy catalogue:
\begin{eqnarray}
i^*_{\rm petro}&<&21,\label{LRG:selection1}\\
0.7(g^*-r^*)+1.2[(r^*-i^*)-0.18] &>& 1.6\,, \label{LRG:scranton1}\\
(g^*-r^*) &>& 1\,, \label{LRG:scranton2} \\
({\rm d}_\perp \equiv) (r^*-i^*)-(g^*-r^*)/8 &>& 0.4\,, \label{LRG:scranton3} \\
r^*_{\rm psf}-r^*_{\rm model}&>&0.24\,,  \label{LRG:selection2}\\ 
\left|z_{\rm phot}-z_{\rm abs}\right|&<&0.05\,. \label{LRG:selection3}
\end{eqnarray}
We also required errors on the model magnitudes to be less than
$0.2{\rm \,mag}$ in $r^*$ and $i^*$, and we excluded objects flagged
by SDSS as BRIGHT, SATURATED, MAYBE\_CR or EDGE. The model magnitudes,
corrected for Galactic extinction, were used to compute the colours.
Equations (\ref{LRG:selection1})--(\ref{LRG:scranton3}) are the LRG
selection criteria of \citet{ScrantonR_04a}. Criterion
(\ref{LRG:scranton3}) is equivalent to imposing $z_{\rm
  phot}\!\ga\!0.3$.  Criterion (\ref{LRG:selection2}) separates stars
from galaxies.  Criterion (\ref{LRG:selection3}) is the selection of
galaxies within a redshift slice of width $\Delta_z\!=\!0.1$ around $z_{\rm
  abs}$ using the photometric redshifts, $z_{\rm phot}$, of
\citet{CsabaiI_03a} who showed that these are accurate to
$\sigma_z\!=\!0.1$.

The choice of the slice width $\Delta_z\!=\!0.1$ corresponds to
$\sim\!200h^{-1}{\rm \,Mpc}$ (co-moving) and is arbitrary. We will
show that the amplitude ratio between the absorber--galaxy
cross-correlation and the galaxy--galaxy auto-correlation does not
depend on the choice of $\Delta_z$ in Section~\ref{section:amplitude} since
we use the same redshift width for both correlation functions.

Finally, we remove the $\sim$10\,per cent of galaxies with problematic
photometric redshifts by requiring that galaxies have $z_{\rm phot}$
uncertainties $\sigma_{z}\!<\!0.5$. A total of \ngalused\ galaxies met
all these criteria in our \nabstot\ fields ($\sim\!800$\,square
degrees).

Figure~\ref{fig:zspec}(right) shows the redshift distribution of these LRGs
for the \nabstot\ fields.  We used the spectroscopic redshift when
available, which includes less than 1\%\ of the sample. This small
fraction is likely to increase in future with the joint 2dF/SDSS
program to obtain spectra of LRGs \citep[e.g.][]{PadmanabhanN_05a}.

\section{Method}
\label{section:method}

We first describe the basics of the galaxy clustering analysis in
Section \ref{section:clustering}. The correlation estimator best used
for this work is discussed in Section \ref{section:estimator}.

\subsection{Galaxy clustering around QSO absorbers}
\label{section:clustering}
 
A widely used statistic to measure the clustering of galaxies is the
correlation function, $\xi(r)$. The absorber--galaxy cross-correlation,
$\xi_{\rm ag}$, is defined from the conditional probability of finding a
galaxy in a volume d$V$ at a distance $ r=|\mathbf r_2\!-\!\mathbf r_1|$,
given that there is a \MgII\ absorber at $\mathbf r_1$:
\begin{equation}
P({\rm LRG}|\MgII) =\overline n_u \left[1+\xi_{\rm ag}(r) \right] \mathrm d
V,
\label{eq:cross}
\end{equation}
where $\overline n_u$ is the unconditional background galaxy density.

Because the observed amplitude of the auto- and cross-correlation
functions are related to the dark matter correlation function,
$\xi_{\rm DM}$, through the mean bias, $\overline b(M)$, which is a
function of the dark matter halo-mass \citep[e.g.][and references therein]{MoH_93a,MoH_02a},
\begin{eqnarray}
\xi_{\rm gg}(r)&=&\overline b^2(M_{\rm g})\,\xi_{\rm DM}(r)\,, \label{eq:biasCDM}\\
\xi_{\rm ag}(r)&=&\overline b(M_{\rm a})\,\overline b(M_{\rm g})\,\xi_{\rm DM}(r)\,, \label{eq:biascross}
\end{eqnarray}
the relative amplitude of the cross-correlation will give a
measurement of the bias ratio and thus of the relative halo-masses:
\begin{eqnarray}
\frac{\xi_{\rm ag}}{\xi_{\rm gg}} = \frac{\overline b(M_{\rm a})}{\overline b(M_{\rm g})} \;. \label{eq:biasratio}
\end{eqnarray}
In other words, if the amplitude of $\xi_{\rm ag}$ is greater
(smaller) than $\xi_{\rm gg}$, the haloes of the absorbers are more
(less) massive than those of the LRGs since the bias increases with
halo mass in all hierarchical models.

In the remainder of this work, we will use only projected correlation
functions, $w(r_\theta)$, where $r_{\theta}=D_A(1+z)\theta$ for
$D_A$ the angular diameter distance in co-moving Mpc.  This is
necessary since our sample is made up of absorbers with spectroscopic
redshifts and of galaxies with photometric redshifts.  The projected
cross-correlation between \MgII\ absorbers and LRGs, $w_{\rm
  ag}(r_\theta)$, is related to $\xi_{\rm ag}(r)$ via $w_{\rm
  ag}(r_\theta)=\int N(z) \xi_{\rm ag}(\sqrt{r_\theta^2+l^2})
\,\mathrm d l$ where $N(z)\equiv\dNdl$ is the line-of-sight
distribution of LRGs and $l$ is the co-moving distance along the
line of sight.

For galaxies distributed along the line of sight as a top--hat function of width $\Delta_z$ (normalized
such that $\int \dNdl \;\mathrm d l=1$), the amplitudes of both
$w_{\rm gg}(r_\theta)$ and $w_{\rm ag}(r_\theta)$ are inversely
proportional to $\Delta_z$ \citep[see also][appendix A]{EisensteinD_03a,BoucheN_05b}:
\begin{eqnarray}
w_{\rm gg}(r_\theta)&=&A_{\rm gg}\;r_\theta^{1-\gamma}\;\simeq\; r_\theta^{1-\gamma}\; r_{\rm 0,gg}^\gamma \; H_\gamma \times
\frac{1}{\Delta_z} \,, \label{eq:wgg}\\  
w_{\rm ag}(r_\theta)&=&A_{\rm ag}\;r_\theta^{1-\gamma}\;\simeq\; r_\theta^{1-\gamma}\; r_{\rm 0,ag}^\gamma \; H_\gamma \times
\frac{1}{\Delta_z}\,, \label{eq:wag}
\end{eqnarray}
where $H_\gamma=\Hgamma$ and $r_{\rm 0,gg}$ \& $r_{\rm 0,ag}$ are, respectively, the
galaxy--galaxy \& absorber--galaxy correlation lengths.
These two equations show that both $w_{\rm ag}$ and $w_{\rm gg}$
depend in exactly the same way on the width of the redshift
distribution $\Delta_z$. \citet[][appendix A]{BoucheN_05b} showed that this
is always true when one correlates one population with known redshifts
(the \MgII\ absorbers) with another population whose redshift is more
uncertain. In Section~\ref{section:amplitude}, we will show
empirically that the ratio $w_{\rm ag}/w_{\rm gg}$ is indeed
independent of $\Delta_z$.  Thus, we stress that $w_{\rm ag}/w_{\rm gg}$ is
{\it independent } of the width of the LRG redshift distribution as
long as one uses the same galaxies for $w_{\rm ag}$ and for $w_{\rm
  gg}$.

From equations~(\ref{eq:wgg}) \& (\ref{eq:wag}), the ratio of the
amplitudes of the two projected correlation functions,
\begin{eqnarray}
a &\equiv & \frac{w_{\rm ag}}{w_{\rm gg}} \,,
\end{eqnarray} 
is simply $(r_{\rm 0,ag}/r_{\rm 0,gg})^{\gamma}$, and is also equal to
the bias-ratio $\overline b(M_{\rm \MgII})/\overline b(M_{\rm LRG})$
[equation~(\ref{eq:biasratio})] from which we infer the mean \MgII\ halo-mass using the bias prescription of \citet{MoH_02a}.

It is important to realise that measuring the halo-mass from {\it the
  ratio} of projected correlation functions has the following
advantages, as advocated in \citet{BoucheN_04a} and
\citet{BoucheN_05b}: (i) one constrains the mass of the \MgII\
host-galaxies in a statistical manner without directly identifying
them, (ii) it is free of possible systematic errors due to
foreground or background contaminants (i.e.~stars or galaxies) and
(iii) it does not require knowledge of the true width of the redshift
distribution of the galaxy population.  The first point is a natural
consequence of correlation statistics. The last two points are
consequences of the fact that the {\it same} galaxies are used to
calculate $w_{\rm gg}(r_\theta)$ and $w_{\rm ag}(r_\theta)$ {\it and}
that both $w_{\rm ag}$ and $w_{\rm gg}$ have the same dependence on
the width of the galaxy redshift distribution
[equations~(\ref{eq:wgg}) \& (\ref{eq:wag})]. This last point follows
from the fact that the absorber redshift is known   precisely
\citep[see ][appendix A]{BoucheN_05b} as mentioned above.  We will
demonstrate points (ii) \& (iii) empirically in
Section~\ref{section:amplitude}.

Using smoothed particle hydrodynamical (SPH) simulations
\citet{BoucheN_05b} showed that the measured amplitude ratio $w_{\rm
  ag}/w_{\rm gg}$ returns the bias-ratio $\overline b_{a}/\overline
b_{g}$ that one expects given the known mass of the simulated
galaxies.  Specifically, the bias formalism of \citet{MoH_02a}
predicts an amplitude ratio of 0.771 in the case of the DLA--galaxy
cross-correlation -- a situation very similar to the \MgII--galaxy
correlation considered here.  Direct measurement of the correlation
functions from the simulations yields $\overline b_{\rm DLA}/\overline
b_{\rm gal}=0.73\pm0.08$, in excellent agreement with the prediction.
The simulations also demonstrated that the cross-correlation technique
does not depend on the galaxy population used to trace the large scale
structure and that foreground and background contaminants do not
effect the measured bias-ratio.

\subsection{Which correlation estimator to use?}
\label{section:estimator}

Which estimator is best used to compute the \MgII--LRG
cross-correlation?  For a given field (i.e.~a single absorber), a
seemingly natural choice would be the estimator
\begin{equation}
1+ w_{\rm ag}(r_\theta) = \frac{\rm G}{\rm  R}\,, \label{eq:estimator:one}
\end{equation}
where G is the observed number of galaxies between $r_\theta-dr/2$ and
$r_\theta+\mathrm d r/2$ around the absorber, and $R$ is the number of
absorber--random galaxy pairs normalized to the total number of
galaxies in the field, $N_{\rm g}$, i.e.~multiplied by $N_{\rm g} /N_{\rm r}$. Ideally,
$R$ would simply be the area of the annulus times the surface density
of galaxies. In practice, some fields are on the edges of the SDSS
coverage, and the SDSS coverage itself is not completely uniform on
the sky; there are small holes, areas missed by SDSS or not yet
released.  One overcomes these problems by (1) generating random galaxies
excluding the gaps and edges, and (2)  by
generating $\sim 200$ times more random galaxies than real galaxies
in order to reduce the shot noise in $R$ to an insignificant proportion.

In our case, we have \nabstot\ fields, at relatively precise
absorption redshifts, $\zabs$.  The selection function for the fields
is therefore a sum of $\delta$-functions.  Should one then use
$\langle \frac{G}{R}\rangle$, where $\langle \rangle$ represents the
averaging operator, in the right-hand side of
equation~(\ref{eq:estimator:one})?  This is clearly not optimum since
it would treat each field equally in performing the average.
\citet{AdelbergerK_03a} (see their equation~B3) showed that a better
choice for the estimator of $w_{\rm ag}(r_\theta)$ is $\langle
G\rangle/\langle R\rangle$, i.e.:
\begin{equation}
1+ {w}_{\rm ag}(r_\theta) = \frac{\rm AG}{\rm AR}\,, \label{eq:estimator}
\end{equation}
where AG\,$=\sum_i G_i$ is the observed number of absorber--galaxy
pairs between $r_\theta-dr/2$ and $r_\theta+\mathrm d r/2$, summed
over all the fields $i$.  AR is the normalized number of
absorber--random galaxy pairs. The normalization is applied to each
field independently: ${\rm AR}\!=\!\sum_i {\rm R}^i \; N_g^i/N_r^i$,
where R$^i$ is the number of random pairs in field $i$, and $N_g^i$
($N_r^i$) is the total number of galaxies (random galaxies) for that
field. Naturally, 
we  took into account the  areas missing from the survey within our search radius. 
Note each absorber's redshift is used to compute $r_\theta$ from a
pair separated by an angular separation $\theta$.

We stress that the \citet{LandyS_93a} estimator is not applicable
here, since it is symmetric under the exchange of the two populations
by one another.  Here, the absorbers have spectroscopic redshifts,
whereas the galaxies have photometric redshifts breaking the symmetry.
Note that this asymmetry has an important consequence: it implies that
both $w_{\rm ag}$ and $w_{\rm gg}$ have the same dependence on the
galaxy redshift distribution, as indicated by equations~(\ref{eq:wgg})
\& (\ref{eq:wag}) \citep[see][appendix A]{BoucheN_05b}.

The cross-correlation computed from equation~(\ref{eq:estimator}) is
biased low due to the integral constraint \cite[e.g.][]{PeeblesP_80a},
explained below. The reason for this is simply that the true
correlation is defined as an overdensity with respect to the
`unconditional' galaxy density $\overline n_u$.  This is not a
measurable quantity as one uses the `observed' galaxy density (or
surface density $\Sigma_g$ for projected correlations) for the field
over which one measures $w(r_\theta)$. As a conseqence, the sum of all
the pairs must be equal to the total number of galaxies $N_g$ observed
in the survey of area $\Sigma$.  This implies that the integral of
$w_{\rm ag}$ over the area $\Sigma$ vanishes: $\int_\Sigma\;w_{\rm
  ag}\;\mathrm d \Sigma=0.$ To fulfill this condition, the
cross-correlation $w_{\rm ag}$ will have to be $<0$ on the largest
scales, i.e.~biased low.  This bias is refered to as the `integral
constraint', $C$.  The true correlation function is then $w_{\rm
  ag}^{\rm C}={ w}_{\rm ag} + C$, where $C$ is the integral
constraint.

There are two ways to estimate $C$. Firstly, it can be estimated
iteratively. For a given amplitude of the correlation, one can
calculate the expected bias $C$ which is then used to correct the
estimate of the amplitude.  The second method is to fit $C$ when one
has sufficient signal to noise.  In the analysis below, we find that
the first method gives $0.0217$, while the second gives
$C=0.018\pm0.012$ (keeping the other parameters fixed).

\section{Results}
\label{section:results}

In this Section, we first present the \MgII--LRG cross-correlation
(Section \ref{section:crosscorr}), and the LRG--LRG auto-correlation
(Section \ref{section:auto}) before showing the main result on the
amplitude ratio between the \MgII--LRG cross-correlation and LRG--LRG
auto-correlation (Section \ref{section:amplitude}).  We turn the
amplitude ratio into a halo-mass for the \MgII\ absorbers in Section
\ref{section:halomass}.  Finally, we show that the amplitude ratio (or
the halo-mass) varies with equivalent width in Section
\ref{section:EW:subsamples}.

\subsection{\MgII--LRG cross-correlation}
\label{section:crosscorr}

\begin{figure}
\centerline{
\includegraphics[width=80mm]{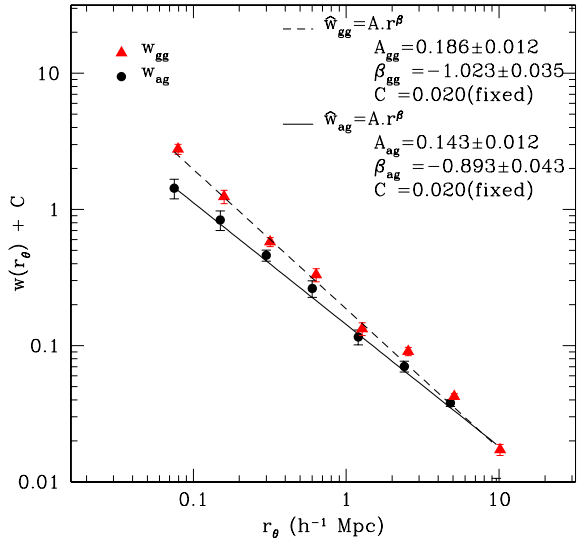}}
\caption{The filled circles show the projected \MgII--LRG
  cross-correlation, $w_{\rm ag}(r_\theta)$, for \nabstot\ absorbers
  and \ngalsim\ LRGs and the triangles show the LRG--LRG
  auto-correlation, $w_{\rm gg}(r_\theta)$.  The error bars are
  computed using the Jackknife technique.  The independent fits to
  $w_{\rm ag}(r_\theta)$ (solid line) and $w_{\rm gg}(r_\theta)$
  (dashed line) are shown.  We fixed the integral constraint $C$ to
  0.02 (see text).}
\label{fig:xcorr:050505:full}
\end{figure}

\begin{table}
\begin{center}
\vspace{-1mm}
\caption{The total number of absorber--LRG pairs, AG, and the number of
  absorber--random pairs, AR, for the cross-correlation shown in
  Fig.~\ref{fig:xcorr:050505:full}.}
\label{table:pairs}
\vspace{-1mm}
\begin{tabular}{ccccc}\hline
	{$r_\theta$} & {AG} & {AR} & {$w_{\rm ag}$} & {$\sigma_{w}$} \\
	{[$h^{-1}{\rm \,Mpc}$]} & { } & { } & { } & { }   \\\hline
 0.05--0.1  &    28        &   11.61   &   1.41    &  0.42	\\
 0.1--0.2   &   82        &   45.13     &   0.82   &  0.19	\\
 0.2--0.4   &   262     &     181.92   &   0.44   &   0.10	\\
 0.4--0.8   &   900       &   724.91   &   0.24    &  0.044 	\\
 0.8--1.6   &   3183      &   2904.27   &   0.096  &  0.015	\\
 1.6--3.2   &   12205    &    11619.1   &   0.050  &  0.010	\\
 3.2--6.4   &   46867     &   46043.9   &   0.018   & 0.007	\\
 6.4--12.8  &   179093    &   181089    &  -0.011   & 0.002    \\\hline
 \end{tabular}
\end{center}
\end{table}

Figure~\ref{fig:xcorr:050505:full} (filled circles) shows the
\MgII--LRG cross-correlation for the entire sample of \nabstot\ \MgII\
absorbers, where we used equation~(\ref{eq:estimator}) for the
estimator of $w_{\rm ag}(r_\theta)$.  There are \ngalused\ objects
within $r_\theta\!=\!12.8h^{-1}{\rm \,Mpc}$ which is the outer radius
of the largest bin used.  Table \ref{table:pairs} shows the total
number of pairs, AG, and the expected number of pairs, AR, if \MgII\
absorbers and LRGs were not correlated.
Figure~\ref{fig:xcorr:050505:full} demonstrates that $w_{\rm
  ag}(r_\theta)$ is a power-law at all scales. As a consequence, the
dip at $\sim 100$\,\hkpc\ in the DR1 sample discussed in
\citet{BoucheN_04a} appears to have been due to small number
statistics.

The error bars for $w_{\rm ag}$ are computed using the jackknife
estimator \citep{EfronB_82a}: we divide the sample into 10 parts and
compute the covariance matrix from the $N_{\rm jack}\!=\!10$
realisations for each part:
\begin{equation}
{\rm COV}_{ij}=\frac{N_{\rm jack}-1}{N_{\rm jack}}\sum_{k=1}^{N_{\rm jack}}  
[w_k(r_{\theta_i})-\overline w(r_{\theta_i})]\cdot [w_k(r_{\theta_j})-\overline w(r_{\theta_j})]
\label{eq:covariance}
\end{equation}
where $w_k$ is the $k$th measurement of the cross-correlation and
$\overline w$ is the average of the $N_{\rm jack}$ measurements.

We fitted $w_{\rm ag}(r_\theta)$ with a power law model, $\hat w_{\rm
  ag}(r_\theta)=A_{\rm ag}\, r_\theta^{\beta_{\rm ag}}$, by minimizing
\begin{equation}
\chi^2\equiv \frac{1}{N_{\rm dof}} [\mathbf{w}-\mathbf{\hat w}]^T \mathrm{COV}^{-1}[\mathbf{w}-\mathbf{\hat w}]\,,
\label{eq:xi}
\end{equation}
where $N_{\rm dof}$ is the number of degrees-of-freedom, $\mathbf{w}$
and $\mathbf{\hat w}$ are the vector data and model respectively, and
${\rm COV}^{-1}$ is the inverse of the covariance matrix. Since ${\rm
  COV}$ is singular we used singular value decomposition techniques to
avoid instabilities in its inversion \citep[see discussion
in][]{BernsteinG_94a}.  Since the integral constraint is
$C=0.02\pm0.01$ (see last Section), we add $\delta_{i,j}(0.01)^2$ to
the covariance matrix ${\rm COV}(w)_{i,j}$ to form ${\rm
  COV}(w^C)_{i,j}$, the covariance matrix for
$w^C(r_\theta)=w(r_\theta)+C$.

Fitting the vector ${\mathbf w}^C$, the best-fit amplitude at
$1h^{-1}{\rm \,Mpc}$ and power-law slope of the cross-correlation are,
respectively,
\begin{eqnarray}
A_{\rm ag}&=&\Across ,  \nn\\
\beta_{\rm ag}&=&\Bcross.\nn
\end{eqnarray}


\begin{figure}
\centerline{\includegraphics[width=80mm]{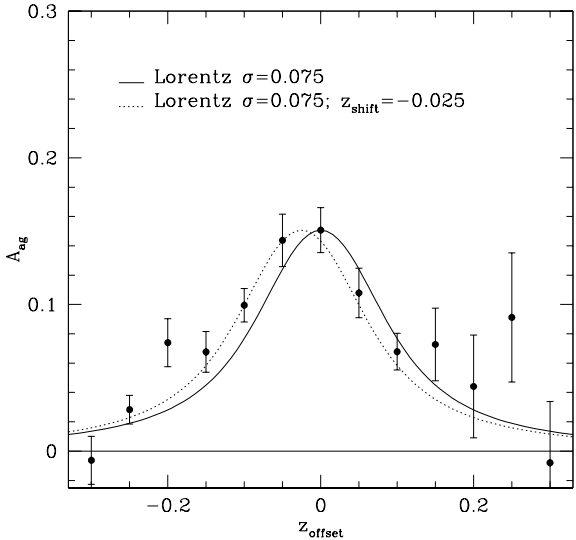}}
\caption{The \MgII-LRG cross-correlation amplitude as a function of
  the artificial redshift offset, $z_{\rm offset}$, imposed on the
  absorber redshifts. As expected, the cross-correlation signal
  vanishes when $|z_{\rm offset}|$ is increased.  This shows that the
  cross-correlation signal in Fig.~\ref{fig:xcorr:050505:full} is
  solely due to the \MgII\ absorbers.  The solid line represents a
  top-hat redshift distribution of width $\Delta_z = 0.1$ convolved with a
  Lorentzian of width ${\rm FWHM}=0.176$, corresponding to the typical
  photometric redshift uncertainty of $\sigma_{z}=0.075$. The solid
  curve is scaled to the $z_{\rm offset}=0$ data point. The dotted
  line shows the solid line shifted by $z_{\rm shift}=-0.025$,
  representing the possible 5\,per cent correction to the photometric
  redshifts suggested by \citet{PadmanabhanN_05a} (see text). The
  increase in the error-bars from left to right is due to the shape of
  the LRG redshift distribution in Fig.~\ref{fig:zspec}(right).}
\label{fig:xcorr:test:z_off}
\end{figure}

\subsection{Luminous red galaxy auto-correlation}
\label{section:auto}

The LRG--LRG auto-correlation $w_{\rm gg}$ is represented by the
grey/red triangles in Fig.~\ref{fig:xcorr:050505:full}.  Since our
goal is to measure the ratio $w_{\rm ag}/w_{\rm gg}$, we are forced to
use the same estimator for $w_{\rm gg}$ as $w_{\rm ag}$, namely
\begin{equation}
1+\overline w_{\rm gg}(r_\theta)= \frac{\rm GG}{\rm GR}\,, \label{eq:estimator:auto}
\end{equation}
where GG is the observed number of galaxy--galaxy pairs between
$r_\theta-\mathrm d r/2$ and $r_\theta+\mathrm d r/2$ and GR is the
number of galaxy--random galaxy pairs, computed as before.  Again,
$r_\theta$ is in units of co-moving Mpc.  The errors and the
covariance matrix for $w_{\rm gg}$ are computed using $N_{\rm
  jack}\!=\!10$ jackknife realisations, as before.  Using the
covariance matrix to $w_{\rm gg}^C$, ${\rm COV}(w^C)_{i,j}$ (as in
Section~4.1), we fitted a power-law, $\hat w_{\rm gg}(r_\theta)=A_{\rm
  gg}\, r_\theta^{\beta_{\rm gg}},$ to $w_{\rm gg}(r_\theta)$.  The
best-fit amplitude at $1h^{-1}{\rm \,Mpc}$ and the power-law slope
are, respectively,
\begin{eqnarray}
A_{\rm gg}&=&\Aauto, \nn\\
\beta_{\rm gg}&=& \Bauto.
\end{eqnarray}

The conversion of the amplitude, $A_{\rm gg}$, to the co-moving
correlation length, $r_{0, \rm gg}$, depends on having precise
knowledge of the true redshift distribution, $N(z)$, of the LRG
sample. The observed distribution is a convolution of the true
distribution and the photometric redshift errors.  Deconvolving these
to find the true distribution is difficult
\citep[e.g.][]{PadmanabhanN_05a} and is beyond the goal of this paper.
However, we note that over a redshift range similar to ours
($z\!=\!0.3$--$0.9$), \citet{BrownM_03a} showed that red
($B_W\!-\!R\!>\!1.44$)\footnote{This cut is similar to ours,
  i.e.~$g-r>1.0$: According to the photometric transformations listed
  on the SDSS website, $B_W-R>1.44$ corresponds to $B-V>1.24$ or
  $g-r>1.14$.} galaxies in the NOAO deep wide survey have a
correlation length of $r_0=6.3 \pm 0.5 h^{-1}{\rm \,Mpc}$, for
galaxies in the luminosity range $-21.5\!<\!M_R\!<\!-20.5$.  At the
mean redshift of our sample, $z=0.5$, such clustering is consistent
with halo-masses of $1\times 10^{13}$\,\msun\ using the bias
prescription of \citet{MoH_02a}.  \citeauthor{BrownM_03a} showed that
the correlation length rapidly increases to $r_{0,\rm
  gg}=11.2h^{-1}{\rm \,Mpc}$ at $M_R\!=\!-22$.  Our sample has a mean
luminosity of $M_{r^\star}\simeq -21.5$, which is consistent with
$r_{0,\rm gg}=8h^{-1}{\rm \,Mpc}$, and a halo-mass slightly higher:
$M_{\rm LRG}=2$--$4\times 10^{13}$\,\msun.  Since we will see that the
systematic errors in our final results are of the same order as the
statistical errors, we hereafter assume\footnote{After
  completion of this analysis and submission of this paper,
  \citet{MandelbaumR_06a} presented a lensing measurement of the shape
  of the density profile of galaxy groups and clusters traced by
  $z<0.3$ LRGs. Their results imply that LRGs fainter than $M_r=-22.6$
  (corresponding to our sample) reside in
  haloes of mass $(2.9\pm0.4)\times10^{13}$~\msun.} that $M_{\rm
  LRG}=3\!\times\!10^{13}$\,\msun\ and treat the uncertainty in its
value as an additional source of systematic errors.  This is
discussed further in Section~\ref{section:halomass}.

\subsection{Is the cross-correlation signal really due to the absorbers?}

In order to verify that the signal of the cross-correlation is solely
due to the \MgII\ absorbers, we repeat our cross-correlation
measurement adding an artificial offset $z_{\rm offset}$ to the
absorber redshift $z_{\rm abs}$, with $z_{\rm offset}$ ranging from
$-0.25$ to $+0.25$.  Here, we fit the cross-correlation $w_{\rm
  ag}(r_\theta)$ using the model $\hat
w(r_\theta)=A\;(r^{\beta}-0.02/0.15)$ in order to account for the fact
that as $A$ tends to zero, so does the integral constraint $C$.  The
slope $\beta $ is kept fixed at $\beta=-0.89$.

Figure~\ref{fig:xcorr:test:z_off} shows the amplitude of the \MgII--LRG
cross-correlation $A_{\rm ag}$ as a function of the artificial
redshift offset $z_{\rm offset}$.  Because the amplitude vanishes when
$|z_{\rm offset}|$ is increased, this figure shows that the \MgII--LRG
cross-correlation signal in Fig.~\ref{fig:xcorr:050505:full} and in
\citet{BoucheN_04a} is solely due to the \MgII\ absorbers.

Given that \citet{CsabaiI_03a} showed that the LRG photometric
redshifts have a typical uncertainty of $\sigma_z\!=\!0.05$--$0.1$, we
convolved our top-hat redshift selection function of width $\Delta_z = 0.1$
with a Lorentzian with ${\rm FWHM}=2.35\,\sigma_z$, representing the
typical photometric redshift uncertainty with an underlying population
of `outliers'.  The solid line in Fig.~\ref{fig:xcorr:test:z_off}
shows the result of the convolution with $\sigma_z=0.075$.  The data
points in this figure might indicate that the photometric redshifts
are slightly over-estimated. In fact, \citet{PadmanabhanN_05a}
showed that at redshifts $z>0.4$, photometric redshifts are slightly
over-estimated by 5\,per cent, i.e.~a redshift correction of
$0.025$ at the mean redshift of our sample. The dotted line shows the
shifted curve.  We emphasize that this will not affect our results
since we measure $w_{\rm ag}/w_{\rm gg}$, which has no dependence on
the redshift distribution of the galaxies.

The error bars in Fig.~\ref{fig:xcorr:test:z_off} increase strongly
with increasing $z_{\rm offset}$. This is due to our redshift
selection function shown in Fig.~\ref{fig:zspec}: as a positive offset
is added, far fewer galaxies are selected and the signal-to-noise
ratio decreases since the number of galaxies directly drives the
number of absorber-galaxy pairs.

\subsection{The relative amplitude $\bmath{w_{\rm ag}/w_{\rm gg}}$}
\label{section:amplitude}

\begin{figure}
\centerline{\includegraphics[width=80mm]{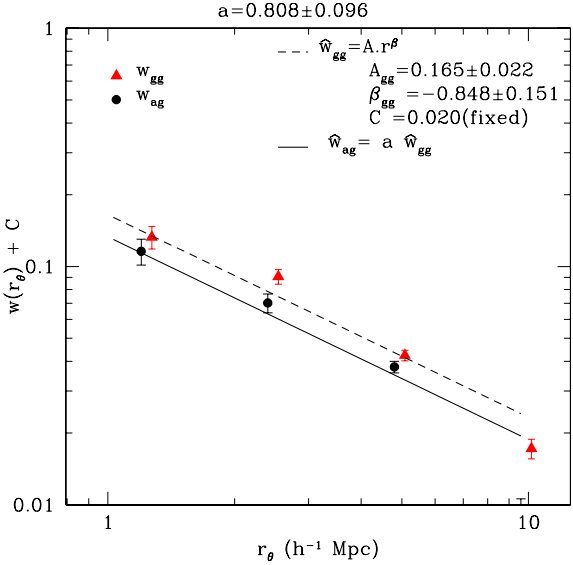} }
\caption{Same as in Fig.~\ref{fig:xcorr:050505:full} in the two-halo
  regime $r_\theta>1$\,\hMpc.  The dashed line represents $\hat w_{\rm
    gg}$ and the relative amplitude of $w_{\rm ag}$ to $w_{\rm gg}$
  (found using $\hat w_{\rm ag}=a\times \hat w_{\rm gg}$) is
  $a=\Arel$. From this, the bias-ratio $b_{\rm \MgII}/b_{\rm LRG}$ is
  $\sim0.65$ (see text), which in turn implies that \MgII\
  host-galaxies have haloes of mass $\Mh\!\simeq10^{12}$\,\msun\ (see
  Section~\ref{section:halomass})}
\label{fig:xcorr:050505:2halo}
\end{figure}

In order to measure the \MgII--LRG bias-ratio [from equations
(\ref{eq:biasCDM}) \& (\ref{eq:biascross})], one needs to use (i) the
scales where the bias $b(M)$ dominates (i.e.~where the correlation
arises from 2 different haloes) and (ii) the same power-law slope,
$\beta$, in order to compare the amplitudes of $w_{\rm ag}$ and
$w_{\rm gg}$.  The first point requires that we use only scales
$r_\theta\!>\!1$\,\hMpc, as numerous papers (both on simulations and on
SDSS data) have shown that the correlations below these scales are
dominated by the single-halo correlation between the central galaxy
and its satellites \citep[e.g.][]{BerlindA_03a,ZehaviI_04a}. The
auto-correlation of $z=0$ SDSS galaxies shows a break at $1$\,\hMpc\
where the transition between the single- and the two-halo terms occur.
Point (ii) is easily achieved by using $\hat w_{\rm gg}$ (Section
\ref{section:auto}) as a template to constrain the relative amplitude
of $w_{\rm ag}$:
\begin{equation}
\hat w_{\rm ag}=a\times \hat w_{\rm gg},
\end{equation}
where $a$ is the amplitude ratio.  

Figure~\ref{fig:xcorr:050505:2halo} shows the auto- and
cross-correlation on scales larger than $r_\theta\!>\!1$\,\hMpc.  We
find that the best relative amplitude is
\begin{equation}
a=\Arel\,. \label{res:arel}
\end{equation}

\begin{figure}
\centerline{\includegraphics[width=80mm]{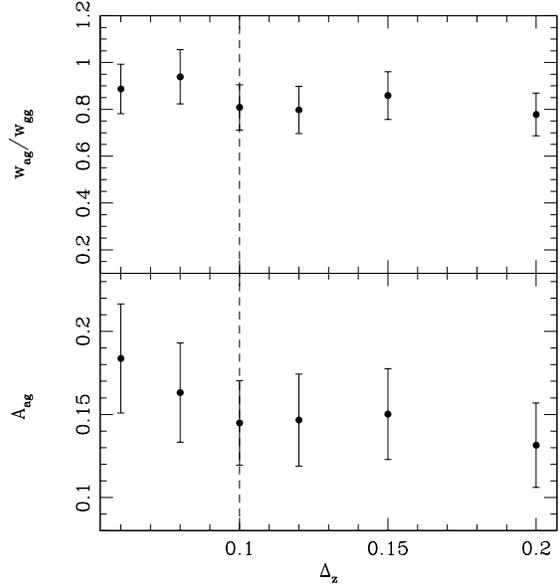}}
\caption{{\it Top}: Amplitude ratio $a$ [equivalently, the bias-ratio
  $\overline b(M_{\rm a})/\overline b(M_{\rm g})$] as a function of
  the width $\Delta_z$ of the redshift slice.  The bias-ratio is
  independent of the redshift width selection.  As a consequence, the
  ratio of the amplitude of the cross-correlation to that of the
  auto-correlation is free of systematics from contaminants (stars or
  interloping galaxies).  {\it Bottom}: The amplitude $A_{\rm ag}$ of
  the cross-correlation $w_{\rm ag}$ decreases as the redshift slice
  selection increases, as expected.  The vertical dashed line shows
  $\Delta_z=0.1$, the selection adopted here
  [equation~(\ref{LRG:selection3})]. Both panels were computed in the  $r_\theta>1$~Mpc regime
  as in   Fig.~\ref{fig:xcorr:050505:full}.}
\label{fig:xcorr:test:Wz}
\end{figure}

As we emphasized in \citet{BoucheN_04a} and pointed out in
Section~\ref{section:crosscorr}, the same galaxies are used to
calculate $w_{\rm ag}$ and $w_{\rm gg}$.  Therefore, the relative
amplitude, $a$, is free of systematics from contaminants (stars or
interloping galaxies).  This is demonstrated in
Fig.~\ref{fig:xcorr:test:Wz}.  The top panel shows $a$ as a function
of the width $\Delta_z$ of the redshift slice.  As one increases $\Delta_z$ up
to $0.2$ ($\simeq 400$~Mpc), i.e.~as one increases the number of
foreground and background galaxies, the amplitude ratio is independent
of that choice.  The bottom panel shows that the amplitude of $w_{\rm
  ag}$, $A_{\rm ag}$, decreases with increasing redshift width, as one
would expect [equation~(\ref{eq:wag})].  However, $A_{\rm ag}$ does
not follow the $1/\Delta_z$ behaviour predicted [\citealt{BoucheN_05b} and
equations~(\ref{eq:wgg}) \& (\ref{eq:wag})].  This is easily explained
given that $\Delta_z$ is a selection criteria upon photometric redshifts:
doubling $\Delta_z$ does not mean we doubled the width of the (true)
redshift distribution since the finite uncertainty in the photometric
redshifts, $\sigma_{z}$, is comparable to $\Delta_z$.

As noted in Section~\ref{section:crosscorr}, the bias-ratio $b_{\rm
  \MgII}/b_{\rm LRG}$ is equal to $a$ in the case of a top-hat
redshift distribution $N(z)$.  However, given the uncertainties in the
photometric redshifts, our LRG sample is distributed around the
absorber redshifts in a Gaussian manner. In \citet{BoucheN_04a} and
\citet{BoucheN_05b}, we showed that, in the case of a Gaussian
redshift distribution, the amplitude of the \MgII--LRG
cross-correlation relative to that of the LRG--LRG auto-correlation,
$a$, is overestimated by \overestimate\,per cent.  
This correction factor was determined using (i) numerical integrations and (ii)
mock catalogues \citep[from the GIF2 collaboration,][]{GaoL_05a}
made of galaxies that had a 
redshift uncertainty equal to the slice width, $\Delta_z$, as in the case 
of our   LRG sample.  Thus, the
bias-ratio inferred from the measured amplitude ratio
equation~(\ref{res:arel})] is
\begin{eqnarray}
\frac{\overline b_{\rm \MgII}}{\overline b_{\rm LRGs}}&=&\Arelcorrsys \,, \label{res:arelsys}
\end{eqnarray}
or \Arelcorrwithsys\ after adding the uncertainties in quadrature.
For comparison, in \citet{BoucheN_04a}, we used 212 \MgII\ absorbers
and $\sim\!20,000$ LRGs selected in SDSS/DR1 and found that the
amplitude ratio was $0.67\pm0.09$.

This bias measurement is entirely consistent with the results of
\citet{BergeronJ_91a} and \citet{SteidelC_94a} who found that \MgII\
absorbers with $\EW\!\geq\!0.3$\,\AA\ are on average associated with
late-type $\sim\!0.7L_B^*$ galaxies, since the expected amplitude
ratio between early and late type galaxies is $\sim\!0.70$
\citep[see][]{BoucheN_04a}.  In addition, the auto-correlation of
late-type galaxies has a shallower power-law slope than that of red
early-type galaxies \citep[e.g.][]{MadgwickD_03a,CollisterA_05a}, a
behaviour recovered in our cross-correlation $w_{\rm ag}$ measurement
in Fig.~\ref{fig:xcorr:050505:full}.

\subsection{Halo-masses of \MgII\ absorbers}
\label{section:halomass}

As already mentioned, in the context of hierarchical galaxy formation,
equations (\ref{eq:biasCDM}) \& (\ref{eq:biascross}) imply that
equation (\ref{res:arelsys}) can be used to infer the halo-mass of the
\MgII\ absorbers provided that the mass of LRGs is known, which is the
case here, as described in Section \ref{section:auto}.

\begin{figure}
\centerline{ 
\includegraphics[width=82mm]{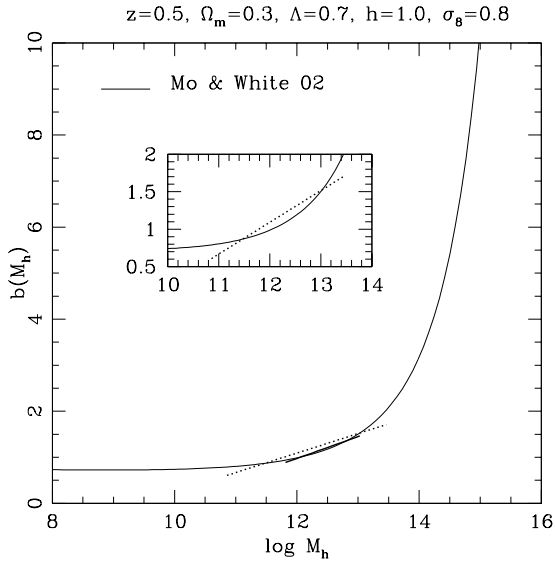}
}
\caption{The solid line shows the bias $b(\Mh)$ as a function of the
  halo-mass $\Mh(\msun)$ using the formalism of \citet{MoH_02a}.  The
  thick solid line shows the linear approximation over the mass range
  $\log \Mh\!=\!$~11.8--13.2 used to estimate that our \MgII\
  absorbers have haloes with mass $\langle \log \Mh (\msun)\rangle
  =\;$\LRGmassrangesys\ [equation~(\ref{eq:mass:MgII})] from the bias
  ratio [equation~(\ref{res:arelsys})].  The dotted line shows the
  linear approximation over the mass range $\log
  \Mh(\msun)\!=\!$~10.7--13.5 used in
  Section~\ref{section:EW:subsamples}.  The inset zooms in on the
  region of interest.  }
\label{fig:bias}
\end{figure}

However, the transformation of equation~(\ref{res:arelsys}) into a
halo-mass is not entirely straightforward.  Indeed, as noted in
\citet{BoucheN_05b}, equations (\ref{eq:biasCDM}) \&
(\ref{eq:biascross}) refer to the mean bias, $\overline b$, averaged
over the mass distributions $\overline b(>\!M)=\int_M^\infty
p(M')\,b(M')\,\mathrm d\log M'\,,$ where $p(M)\equiv \frac{\mathrm d
  n}{\mathrm d\log M}$ is the appropriate mass distribution
[normalized such that $\int p(M)\,\mathrm d\log M=1$] and $b(M)$ is the
bias of haloes of a given mass $M$, shown in Fig.~\ref{fig:bias}.  In
the case of the LRGs, $p(M)$ is given by the halo mass function
$n(M)$.  Unfortunately, the distribution $p(M)\equiv{\mathrm d
  n}{/\mathrm d z/\mathrm d\log M}$ for the \MgII\ absorbers is unknown.

As discussed in \citet{BoucheN_05b}, this issue may be alleviated by
expanding $b(M)$ to first order over a restricted mass range, i.e.,
$b(M)=b_0+b_1\;\log M$. The mean bias $\overline b$ is then given by
\begin{eqnarray}
  \overline b(>\!M_{\rm min})&=&\int_{M_{\rm min}}^\infty p(M')\,b(M')\,\mathrm d\log M'  \nn \\
  &=&b_0+b_1\,\int_{M_{\rm min}}^\infty p(M')\, \log M'\,\mathrm d\log M' \nn \\
  &=&b_0+b_1\,\langle\log M\rangle  \label{eq:bias:approx}\,,
\end{eqnarray} 
where $\langle \log M\rangle $ is the first moment of the distribution
$p(M)$.  For example, the mean bias of the \MgII\ absorbers is
$\overline b$ evaluated at the mean halo-mass, i.e.~$\overline b_{\rm
  \MgII}=\overline b(\langle\log \Mh\rangle)$, where $<>$ is the
average using the appropriate mass distribution $p(M)$.  Since the
coefficients $b_0$ \& $b_1$ can be obtained from the bias function
$b(M)$, the distribution $p(M)$ need not to be known.

The thick dashed line in Fig.~\ref{fig:bias} shows the linear fit to
$b(M)$ over $\sim 1.4$~dex.  Provided that the mass range of interest
is small ($\la 1$~dex), this approximation is valid.  Using this
linear fit to $b(M)$, our bias-ratio measurement
equation~(\ref{res:arelsys}) implies that the \MgII\ absorbers reside
in haloes of mean mass $$\langle \log \Mh (\msun)\rangle
=\LRGmassrangesysnolrg,$$ regardless of the actual $p(M)$ distributions.
An additional source of systematic error is
the mass of LRGs, which has been assumed to be
$\sim\!3\!\times\!10^{13}$\,\msun.  Given that the uncertainty in the
correlation length $r_{0, \rm gg}$ is of the order of $\sim\!1$~Mpc at
most (Section~\ref{section:auto}), we find that the uncertainty in the
LRG halo-mass is $^{+0.21}_{-0.25}$~dex.  This implies that the
additional systematic uncertainty in the absorber halo-mass is
$^{+0.14}_{-0.16}$~dex.
 
By adding this additional systematic uncertainty in quadrature, we find that
the mean halo-mass of \MgII\ absorbers is (with 1-$\sigma$ uncertainties):
\begin{equation}
\langle \log \Mh (\msun)\rangle =\LRGmassrangeallsys\,. \label{eq:mass:MgII}
\end{equation}
 Adding the statistical
and systematic errors in quadrature gives $\langle \log \Mh
(\msun)\rangle =\LRGmassrangesys$.

Is this mass scale $\sim\!10^{12}$\,\msun\ consistent with the
incidence probability of \MgII\ absorbers? In other words, are there
too many (or too few) such haloes in a $\Lambda$CDM universe?  The
incidence probability of \MgII\ absorbers, \dNdz, is given by the
volume number-density of haloes, $n$, times the co-moving
cross-section, $\sigma_{\rm co}$:
\begin{equation}
\frac{\mathrm d N}{\mathrm d z} \;=\;  n\,\sigma_{\rm co}\,\frac{\mathrm d r}{\mathrm d z}\,, \label{eq:dNdz}
\end{equation}
where the cross-section $\sigma_{\rm co} = \pi R_{\rm
  phys}^2(1+z_{\rm abs})^2$ for $R_{\rm phys}$ the radius of the
cross-section in physical units.  Since $R_{\rm phys}$ is
$\sim$40\,\hkpc\ \citep{SteidelC_95b} and the number density of haloes
of mass $M=10^{12}$\,\msun\ is $n=10^{-2}h^{-3}$~Mpc$^{-3}$
\citep{MoH_02a}, we find that $\dNdz\simeq0.3\;(n/10^{-2})\;(R_{\rm
  phys}/40\,\hbox{\hkpc})^2$.  This is close to the observed value for
\MgII\ absorbers with $\EW\ga 1$\,\AA\
\citep{NestorD_05a,ProchterG_05a}. We can therefore conclude that
there are neither too few nor too many $\Mh=10^{12}$\,\msun\ haloes to
account for the observed incidence probability.

\subsection{Are $\bmath{W_{\rm r}}$ and $\bmath{M_{\rm h}}$
  anti-correlated?}
\label{section:EW:subsamples}

\begin{table}
\begin{center}
  \caption{The bias-ratio and the inferred halo-mass of \MgII\
    absorbers, with 1-$\sigma$ statistical uncertainties, as a
    function of \EW, binned in two different ways. The first splits
    the sample of nabstot\ \MgII\ absorbers into 4 bins of equal
    size except for the last bin which is double the size to provide
    enough statistics. The second uses 3 bins of equal size.
    Regardless of the binning, the bias-ratio decreases with
    increasing \EW.
    \label{table:biasratios}}
\vspace{-5mm}
\begin{tabular}{lccc}
\hline
\EW  (\AA) 	 &  $a\equiv w_{\rm ag}/w_{\rm gg}$   &$b_{\rm \MgII}/b_{\rm LRG}$   &    $\log M_{\MgII}(\msun)$ \\
\hline
0.3--5.5	 & \Arel	&  \Arelcorr 	&  \LRGmassrange \\\hline
0.30--1.15	 &  0.99   $\pm$   0.11  &   0.79 $\pm$     0.09  &  $12.49\pm0.34$  \\
1.15--2.00	 &   0.81   $\pm$    0.11  &  0.65 $\pm$    0.08  &  $11.93\pm0.33$ \\
2.00--2.85	 &   0.54   $\pm$   0.17   &  0.43 $\pm$    0.13  &  $11.11^{+0.52}$ \\
2.85--5.50	 &   0.66   $\pm$    0.24  &  0.53 $\pm$    0.19  &  $11.47^{+0.74}$ \\\hline
0.725--1.575	 & 1.13        $\pm$	0.12	&  0.89       $\pm$	0.09 	&  $12.88\pm0.37$  \\
1.575--2.415	 & 0.70       $\pm$	0.13  &    0.56      $\pm$	0.11	 &  $11.59\pm0.41$  \\
2.415--4.115	 & 0.61       $\pm$	0.14	&  0.49      $\pm$	0.11  	&   $11.30\pm0.43$ \\
\hline
\end{tabular}
\end{center}
\end{table}

\begin{figure*}
\centerline{
\includegraphics[width=80mm]{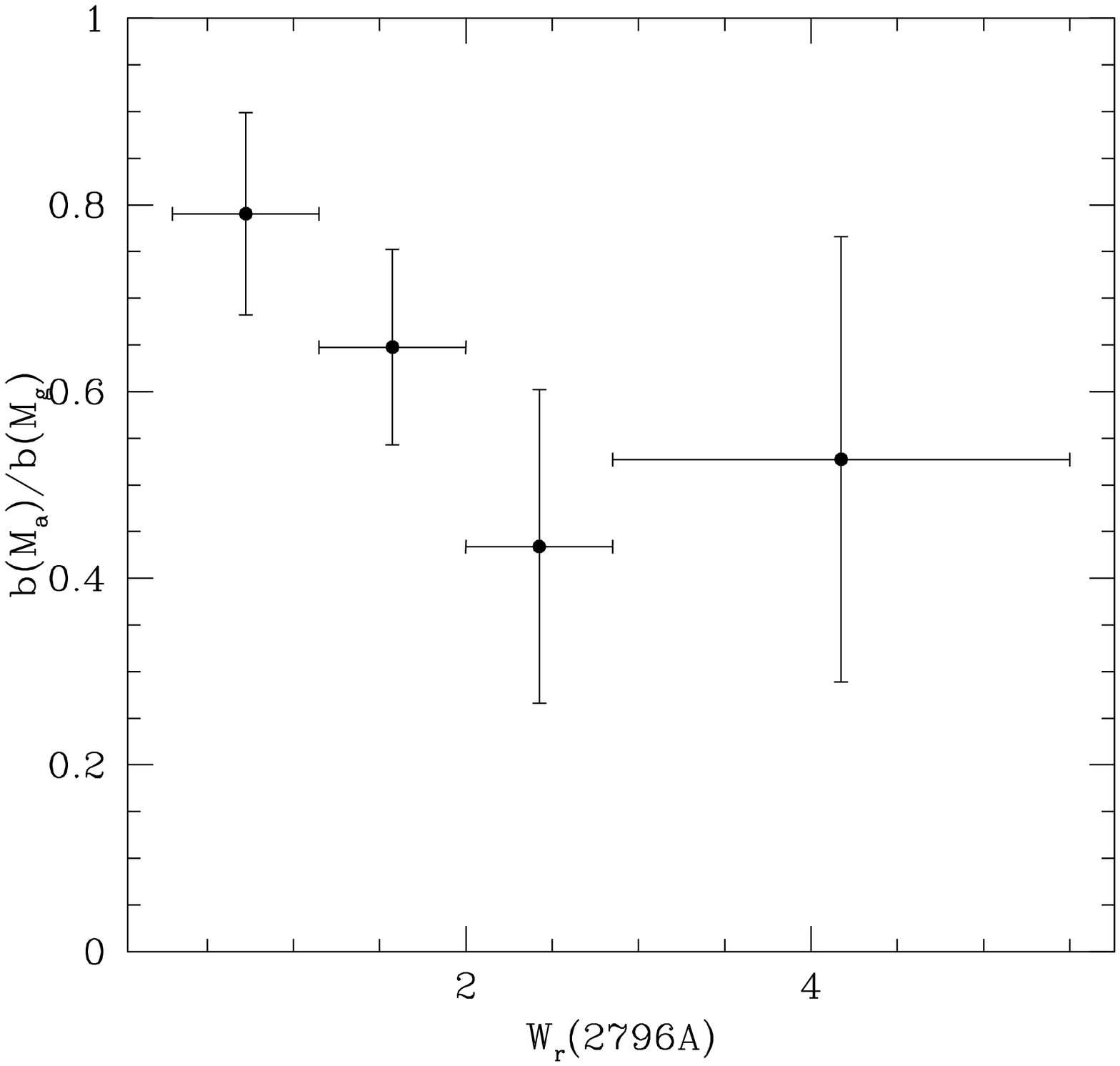}
\includegraphics[width=80mm]{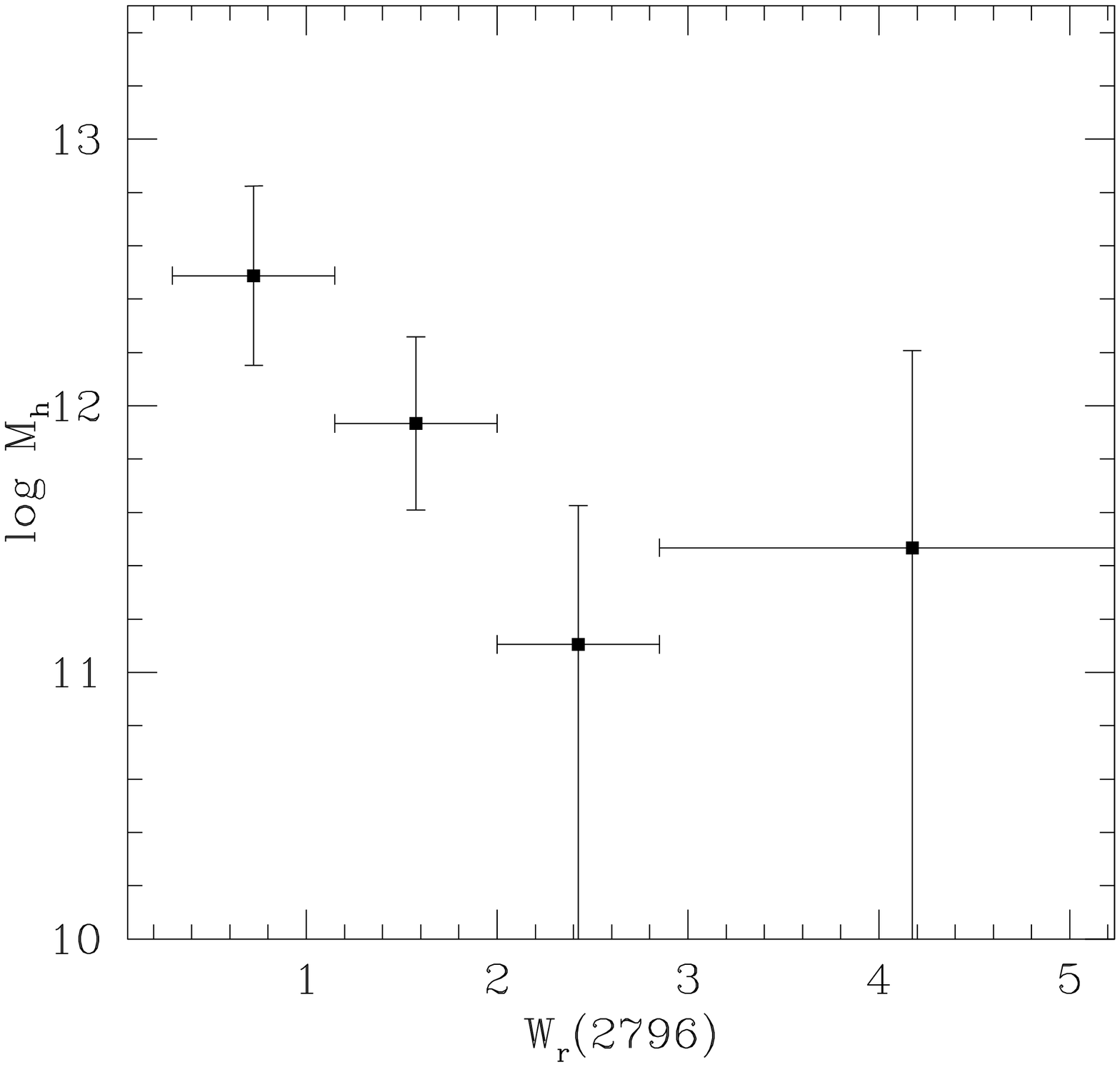}
}
\caption{{\it Left}: The bias-ratio $b(M_{\rm a})/b(M_{\rm g})$ as a
  function of rest-frame equivalent width \EW\ using 3 equally spaced
  bins plus a larger bin. The bias-ratio is computed from the ratio
  of the \MgII-LRG cross-correlation to the LRG-LRG auto-correlation
  measured on scales $1$--13~\hMpc\ as in Section~4.4.
  The bias-ratio is anti-correlated with \EW.
  This anti-correlation implies that
  higher equivalent width absorbers are less massive. 
  {\it Right}: The halo-mass,
  $\Mh(\msun)$, of \MgII\ absorbers as a function of \EW\ in bins
  corresponding to those on the left plot.  Absorbers with
  $\EW\!\ga\!1.5$\,\AA\ have a halo-mass 1\,dex smaller than absorbers
  with $0.3\!\la\!\EW\!\la\!1.0$\,\AA. }
\label{fig:xcorr:EW}
\end{figure*}

We searched for possible correlations between the halo-masses of our
\nabstot\ \MgII\ absorbers and various parameters of our survey such
as the absorber redshift $z_{\rm abs}$, the quasar redshift $z_{\rm
  qso}$, \MgII\ equivalent width, signal-to-noise of the SDSS spectra,
\EW, the quasar magnitude and the quasar colours. We did not find any
significant correlation with these parameters except for the
equivalent width, \EW. Our DR1 sample \citep{BoucheN_04a} already
showed tentative evidence that the cross-correlation had a slightly
smaller amplitude for higher equivalent width absorbers. Here we
significantly strengthen that evidence.

Figure~\ref{fig:xcorr:EW}(left) shows the bias-ratio $b_{\rm
  \MgII}/b_{\rm LRG}$ for sub-samples in 4 equivalent width bins of
equal width except for the last bin which is twice as large in order
to have enough statistics. This figure shows that the bias-ratio, and
therefore the halo-mass, is anti-correlated with \EW. We also divided
our sample into 3 sub-samples of 602 absorbers.
Table~\ref{table:biasratios} lists the halo-masses for each \EW\ bin
and shows that the anti-correlation is robust under different binning.
The bias-ratio in this figure is computed from the ratio of the
\MgII--LRG cross-correlation to the LRG--LRG auto-correlation measured
on large scales ($1$--13~\hMpc) as in Section~4.4.  In other words,
the anti-correlation shown in Fig.~8 has nothing to do with `\MgII\ absorbers in
the haloes of LRGs', since the haloes of LRGs
are much smaller than the scales probed by our clustering.

Figure~\ref{fig:xcorr:EW}(right) shows the halo-mass inferred for each
of the 4 bins used in the left plot.  In converting bias-ratio to
halo-mass, we used the same method outlined in Section
\ref{section:halomass}.  However, since the bias-ratio in
Fig.~\ref{fig:xcorr:EW}(left) corresponds to a larger mass range than
in Section \ref{section:halomass}, we adopted the dotted line shown in
Fig.~\ref{fig:bias} as a linear approximation to the bias function
[equation~(\ref{eq:bias:approx})].  This approximation is naturally
poorer than for our mass estimate for the entire sample
[equation~(\ref{eq:mass:MgII})].  Of course, equation
(\ref{eq:bias:approx}) can be expanded to the second order, but it
would include the second moment $\langle(\log M)^2\rangle$ of the
distribution $p(M)$, which is unknown. Quantitatively, we find that
absorbers with $W_{\rm r}^{\ma}\ga 2$\,\AA\ have haloes with $\langle
\log M_{h} (\msun)\rangle=\;$\LRGmassrangelow, while \MgII\ absorbers
with $0.3\!\la \! \EW \!\la 1.0$\,\AA\ have $\langle \log M_{h}
(\msun)\rangle=\;$\LRGmassrangehigh, i.e.  a mass difference of
$\ga$1.0~dex.

The observed $M_{\rm h}$--\EW\ anti-correlation in
Fig.~\ref{fig:xcorr:EW} has important implications. If \MgII\ clouds
were virialized entities in the gaseous haloes of the host-galaxies,
one would expect that the velocity spread of \MgII\ systems would be
proportional to the mass of the host galaxies. At equivalent widths
$\EW\!\ge\!0.3$\,\AA, the \MgII\ transitions of the sub-components are
saturated and so \EW\ strongly correlates with the velocity spread
($\Delta v$) of the absorber \citep[e.g.][]{EllisonS_06a}.
This velocity range for strong \MgII\ systems
 covers the range 50--400~\kms. 
 If the individual clouds are virialized in the halo of the host-galaxies,
$\Delta v$ ought to be related to the velocity dispersion of the
gaseous halo and to the mass of the host galaxy. Therefore, the
$M_{\rm h}$--\EW\ anti-correlation shows that most strong \MgII\
absorbers are {\it not} virialized in the gaseous haloes of their host-galaxies.

In fact, it is often tacitly assumed in the literature that $M_{\rm
  h}$ and $\Delta v$ (as traced by \EW) should be positively
correlated and so the $M_{\rm h}$--\EW\ anti-correlation in
Fig.~\ref{fig:xcorr:EW} may seem surprising at first. In the next
Section, we show that this auto-correlation is compatible with numerous
past results in the literature.  Moreover, those past results actually
seem to {\it require} a $M_{\rm h}$--\EW\ anti-correlation, independent
of our new results.

Recently, \citet{ProchterG_05a} used the redshift evolution of \dNdz\
and simple cross-section arguments to infer rough estimates of the
halo-mass of \MgII\ absorbers.  Their conclusions are similar to ours:
absorbers with large equivalent width, $\EW\ga1.8$\,\AA, were found to
be more likely associated with $\Mh<M_\star$ haloes, not with massive
$M_\star\sim10^{12}$\,\msun\ haloes. 

\section{Do past results require a $\bmath{\Mh}$--$\bmath{\EW}$
  anti-correlation?}
\label{section:discussion}

In this Section, we study the incidence probability of absorbers in
more detail.  Specifically, we use two arguments to show that a
$M_{\rm h}$--\EW\ anti-correlation is already required by past
results, i.e.  regardless of our cross-correlation results.  In
Section \ref{section:discussion:cross}, we show that the
cross-section--luminosity relationship combined with the
cross-section--equivalent width relationship require an
anti-correlation between luminosity and equivalent width.  In Section
\ref{section:discussion:lum}, we show that the
cross-section--luminosity relationship, the observed incident
probability (\dNdzdW) and any plausible host-galaxy luminosity
function (LF) together require an anti-correlation between luminosity
and equivalent width. We also show that the slope of the expected
anti-correlation is completely consistent with our new result in
Fig.~\ref{fig:xcorr:EW}.

The starting point for the discussions below is the incidence
equation, equation~(\ref{eq:dNdz}), but in its integral form:
\begin{eqnarray}
&&\int_{W_1}^{W_2}\frac{\mathrm d^2 N}{\mathrm d W_{\rm r} \mathrm d z}(W_{\rm r})\mathrm d W_{\rm r} \nonumber\\
&=& \int_{M_1}^{M_2}\mathrm d\log M\, \frac{\mathrm d^2 N }{\mathrm d \log M \mathrm d z}(M) \nonumber\\
&=& \int_{M_1}^{M_2}\mathrm d\log M\, \frac{\mathrm d^2 N }{\mathrm d \log M \mathrm d V}(M) \,\sigma_{\rm co}(M)\,
\frac{\mathrm d r}{\mathrm d z}\,,\label{eq:dNdz:integral}
\end{eqnarray}
where ${\mathrm d^2 N }/{\mathrm d \log\!M /\mathrm d V}$ is the
number of haloes (or galaxies) per unit volume per unit mass (or
luminosity).  In addition, a covering factor $C_{\rm f}$ should be
included in the right-hand side of this equation since observations
imply that it may be less than unity: \citet{TrippT_05a} studied close
galaxy--QSO pairs and found that $\sim$50\,per cent of galaxies do not
produce any \MgII\ absorption down to 0.3\,\AA, indicating that the
mean covering factor for our sample is $C_{\rm f}\sim 0.5$. We note
that $C_{\rm f}$ is likely to depend on the size of the host galaxy
and therefore its mass, a complication that we ignore here.

For the past two decades, the main problem in solving equation
(\ref{eq:dNdz:integral}) for the cross-section $\sigma_{\rm co}$ was
that one had to perform the integrals of the left-hand side over a
range of \EW\ and the right-hand side over a range of mass $M$ (or
luminosity $L$), ignoring any possible -- and probably quite strong --
dependence of \EW\ on $L$ or $M$.  We will show in the next two
sub-sections that the $M_{\rm h}$--\EW\ anti-correlation is required
(independently of our cross-correlation results).

We first show (in Section 5.1) that a $M_{\rm h}$--\EW\ relationship
is required from the following simple arguments. Observationally, for
a given bin in \EW, there is a maximum impact parameter, $R_{\rm
  phys}$, that defines the cross-section, $\sigma$. The radius of the
cross-section depends on the luminosity of the host galaxy, $R_{\rm
  phys}\propto L^{\beta}$, i.e.~the cross-section is a function of
mass or luminosity.  Since there is a relationship between \EW\ and
$R_{\rm phys}$ and between $R_{\rm phys}$ and $L$, there is a
relationship between \EW\ and \Mh\ (or $L$). Section~5.1 shows
that the two physical parameters should be anti-correlated.
Section~5.2 arrives at the same conclusion based on different
arguments which make one assumption, namely that the relationship
between $\log \Mh$ and \EW\ is linear.

\subsection{The cross-section distributions require a
  $\bmath{\Mh}$--$\bmath{\EW}$ anti-correlation}
\label{section:discussion:cross}

\begin{figure}
\centerline{
\includegraphics[width=80mm]{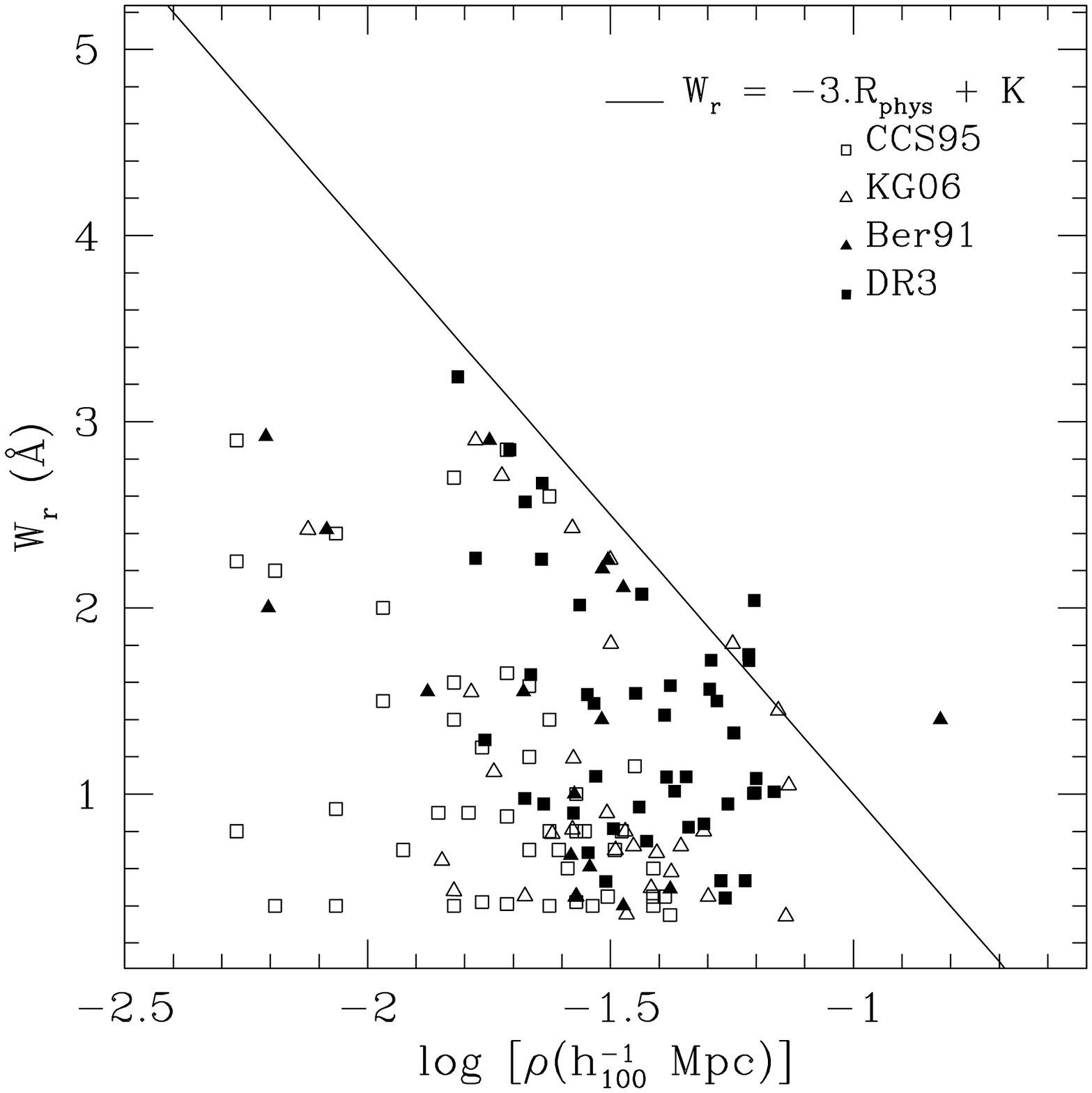}
}
\caption{The \MgII\ equivalent width, \EW, as a function of impact
  parameter, $\rho$, for the host-galaxies of absorbers with
  $\EW\ga0.3$\,\AA.  The solid squares show the impact parameters for
  host-galaxies selected from a sample of our SDSS \MgII\ absorbers
  closer than 100\hkpc\ (co-moving) to the QSO lines-of-sight and with
  $|z_{\rm phot}-z_{\rm abs}|<0.05$. The open squares, open triangles
  and solid triangle show the samples of \citet{SteidelC_95b},
  \citet{KacprzakG_06a} and \citet{BergeronJ_91a} respectively. All
  impact parameters are shown in our cosmology.  The solid line shows
  the fiducial model for the maximum impact parameter $R_{\rm
    phys}$ used in Section~\ref{section:discussion:cross} (see text)
  to represent the cross-section--equivalent width relationship,
  $\sigma_{\rm co}(\EW)$.}
\label{fig:impact-EW}
\end{figure}

Here, we investigate the existence of a luminosity (or mass)--\EW\
relationship and whether the two parameters should be correlated or
anti-correlated given the observed distribution of \MgII\ absorbers in
the cross-section--equivalent width plane and the dependence of the
cross-section on luminosity \citep{SteidelC_95b}.

The radius of the cross-section, $R_{\rm phys}$, decreases steeply
with increasing \EW: \citet{LanzettaK_90a} and \citet{SteidelC_95b}
(their figure~3) showed that the host-galaxies of absorbers with
$\EW>2$\,\AA\ do not exist at large impact parameters.
Figure~\ref{fig:impact-EW} reproduces these results.  The data points
show the impact parameters $\rho$ (in physical units) of the \MgII\ host
galaxies from the samples of \citet{BergeronJ_91a} (filled triangle), 
\citet{SteidelC_95b} (open squares) and
\citet{KacprzakG_06a} (open triangles) for absorbers with
$\EW>0.3$\,\AA.  The solid squares represent the 44 host-galaxies
that we have identified in the SDSS/DR3 imaging data with $|z_{\rm
  abs}-z_{\rm phot}|<0.05$ and which were within 100~kpc (co-moving)
of the QSO line-of-sight. This sample is truncated at an impact
parameter $\rho\la 20$\,\hkpc\ (physical; $\sim$3\arcsec) due to the large
point-spread function of SDSS.  It is clear that the data points do
not fill the upper right side of the plot. More precisely, at
increasing equivalent width, the maximum impact parameter -- which
defines the radius $R_{\rm phys}$ of the cross-section --
decreases. This dependence can be roughly parametrized as
$W_{\rm r}(R_{\rm phys}) \simeq -3 \log [R_{\rm phys}]+K$ as
represented by the solid line in Fig.~\ref{fig:impact-EW}.

In addition, a relationship between between $R_{\rm phys}$ and
luminosity $L$ also exists: \citet{SteidelC_95b} found that $R_{\rm
  phys}$ slightly increases with luminosity: $R_{\rm
  phys}(L)\propto (L_K/L_K^*)^{\beta}$, with $\beta\simeq0.2$.  Thus,
the simple observational results that there are relationships between
$W_{\rm r}$ \& $R_{\rm phys}$ and between $R_{\rm phys}$ \& $L$
require a $L$--\EW\ relationship.

We emphasize that this $L$--\EW\ relationship does not mean that there
is a direct correspondence between mass and equivalent width on a
galaxy-by-galaxy basis. This relationship simply means that {\it on
  average} a sample with a well-defined \EW\ (e.g.~a bin of \EW) will
have a well-defined and predictable mean mass (or luminosity). More
specifically, for a sample of absorbers selected in a bin of \EW, the
host galaxy will have an impact parameter less than $R_{\rm phys}$,
and a mean mass \Mh.  One should not interpret the distribution in
\EW\ as being entirely due to the distribution in halo mass (or in
galaxy luminosity).

Finally, if one combines the observed cross-section--equivalent width
relation (Fig.~\ref{fig:impact-EW}) with this
cross-section--luminosity relation one finds that $W_{\rm r}^{\rm
  phys}(\log L_K)\propto -3\beta\log L_K\simeq -0.6 \log L_K$,
i.e.~$\log L_K \propto \frac{1}{-3 \beta}W_{\rm r} \simeq -1.6 W_{\rm
  r}$.  In other words, the cross-section as a function of luminosity
and as a function of equivalent width {\it together require} that
$W_{\rm r}$ (therefore $\Delta v$) and luminosity (or mass) be
anti-correlated.

\subsection{The incidence probability requires a
  $\bmath{\Mh}$--$\bmath{\EW}$ anti-correlation}
\label{section:discussion:lum}

\begin{figure}
\centerline{ 
\includegraphics[width=7cm]{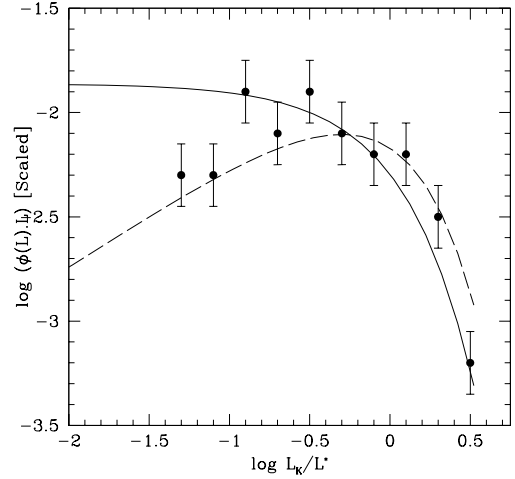}
}
\caption{The symbols represent the absorber $K$-band luminosity
  function (LF) of \citet{SteidelC_95b}. The error bars are uniform
  and represent the average uncertainties.  The solid line represents
  a Schechter function with $\alpha=-1$, as observed for the field
  $K$-band LF \citep{KochanekC_01a}.  The dashed line represents a
  Schechter function with $\alpha=-0.5$.  Both lines are reasonable
  descriptions of the $K$-band LF.  Both curves were scaled to match
  the data points.  }\label{fig:LF}
\end{figure}

Since we have established the existence of a \Mh--\EW\ relationship,
the right-hand side of equation (\ref{eq:dNdz:integral}) can be
rewritten as
\begin{eqnarray}
  && \int_{W_1}^{W_2}\mathrm d W_{\rm r} \frac{\mathrm d\log M}{\mathrm d W_{\rm r}}\, \frac{\mathrm d^2 N }{\mathrm d \log M \mathrm d V}
  \,\sigma_{\rm co}\,\frac{\mathrm d r}{\mathrm d z}\,.\nonumber 
\end{eqnarray}
By differentiation of both sides of equation (\ref{eq:dNdz:integral}),
it follows that~\footnote{If $F(x)\equiv\int^x f(a) \mathrm d a$ and
  $G(x)\equiv\int^x g(a) \mathrm d a$. If $F(x)=G(x)$ for all $x$ then
  $F=G$ and their derivatives are also equal, $f(a)=g(a)$.}
\begin{eqnarray}
  \frac{\mathrm d^2 N}{\mathrm d W_{\rm r} \mathrm d z}
  &=&  \frac{\mathrm d\log M}{\mathrm d W_{\rm r}}\, \frac{\mathrm d^2 N }{\mathrm d \log M \mathrm d V}
  \,\sigma_{\rm co}\,\frac{\mathrm d r}{\mathrm d z}\,,\label{eq:dNdz:diff}
\end{eqnarray}
where $\frac{\mathrm d\log M}{\mathrm d W_{\rm r}}$ is the Jacobian of
the transformation between $W_{\rm r}$ and $M$. We will assume that
the relationship between $W_{\rm r}$ and $M$ is linear,
i.e.~$\frac{\mathrm d\log M}{\mathrm d W_{\rm r}}$ is a constant.

An independent way to show the existence of a $L$--$W_{\rm r}$
anti-correlation makes use of equation~(\ref{eq:dNdz:diff}) in terms
of luminosity:
\begin{eqnarray}
\frac{\mathrm d^2 N}{\mathrm d z\mathrm d W_{\rm r}}  
&=& \frac{\mathrm d\log L}{\mathrm d W_{\rm r}}\, \frac{\mathrm d^2 N}{\mathrm d V\mathrm d \log L}\;\sigma_{\rm co}(L)\; \frac{\rm d r}{\rm d z} \label{eq:dNdzdlogL} \,,
\end{eqnarray}
where ${\mathrm d^2 N}/{\mathrm d V/ \mathrm d \log\!L}$ is simply the
LF in logarithmic form, $\phi(L)L\ln(10)$.  The differential incidence
probability, \dNdzdW, is now well constrained observationally thanks
to the large SDSS database \citep[e.g.][]{NestorD_05a,ProchterG_05a}.

For luminosities below $L_*$, $\phi(L)L\propto L^{1+\alpha}$, where
$\alpha$ is the faint-end power-law slope of the LF.
Equation~(\ref{eq:dNdzdlogL}) then implies that
\begin{eqnarray}
 \frac{\mathrm d^2 N}{\mathrm d z\mathrm d \log L} &\propto&L^{(1+\alpha+2\beta)} \,\label{eq:dNdzdlogL:approx}
\end{eqnarray}
using $\sigma_{\rm co}(L)\propto L^{2\beta}$ from \citet{SteidelC_95b}. In
addition, \citet{NestorD_05a} showed that \dNdzdW\ is an exponential,
i.e.
\begin{eqnarray}
 \frac{\mathrm d^2 N}{\mathrm d z\mathrm d W_{\rm r}}  &\propto &  \exp(-W_{\rm r}/W_*)\,,\label{eq:dNdz:approx}
\end{eqnarray}
where $W_*$ is the exponential scale length. 

Quantitatively, from equations~(\ref{eq:dNdzdlogL:approx}) and
(\ref{eq:dNdz:approx}), the expected slope of the $L$--$W_{\rm r}$
anti-correlation is
\begin{eqnarray}
\frac{\rm d\log L}{\rm d W_{\rm r}}& \simeq& \frac{-\log_{10}(e)/W^*}{1+\alpha+2\beta},\\
&\simeq&-1.8\; \hbox{for $\alpha=-1.0$, or}\nn\\
&\simeq&-0.8\; \hbox{for $\alpha=-0.5$}\nn,
\end{eqnarray}
since \citet{NestorD_05a} showed that $W^*\sim0.6$ in the redshift
range $\langle z\rangle\sim 0.5$ of our survey, and
\citet{SteidelC_95b} showed that $\beta=0.2$ .

Therefore, equations~(\ref{eq:dNdzdlogL:approx}) and
(\ref{eq:dNdz:approx}) imply that $W_{\rm r}$ and $\log L$ (or $\log
M$, using a constant $M/L$) must be anti-correlated as long as
${1+\alpha+2\beta}>0$, i.e.~$\alpha>-1.4$ for $\beta=0.2$. Most
plausible LFs easily satisfy this condition.

The $B$-band LF of \MgII\ absorbers was constrained by
\citet{SteidelC_95a} and is known to be bell-shaped, with $\alpha$
shallower than $-1$ \citep[figure~4 of][]{SteidelC_95b}.  The $K$-band
LF of \MgII\ is likely to be closer to a mass function.
Figure~\ref{fig:LF} reproduces the absorber $K$-band luminosity
function of \citet{SteidelC_95b}. The error bars are uniform and
represent the average uncertainties. The solid line represents a
Schechter function with $\alpha=-1$, as observed for the $z=0$
$K$-band field-galaxy LF \citep{KochanekC_01a,BellE_03a}.
\citet{SteidelC_95b} argued that the host-galaxy $K$-band luminosity
function is similar to the $K$-band field-galaxy LF. The dashed line
represents a Schechter function with $\alpha=-0.5$ and is an equally
good match to the data.

\begin{figure*}
\centerline{
\includegraphics[width=80mm]{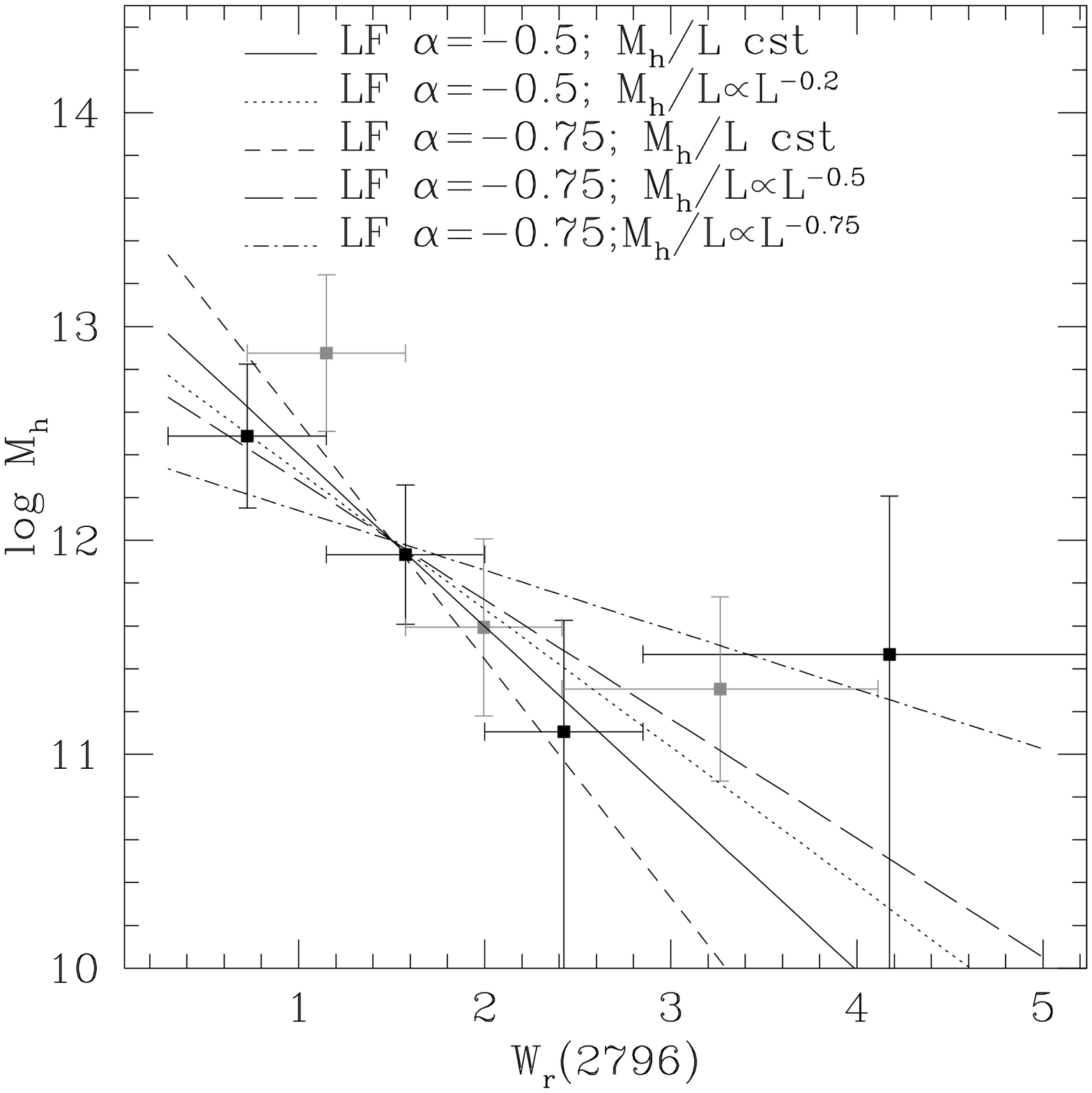}
\includegraphics[width=80mm]{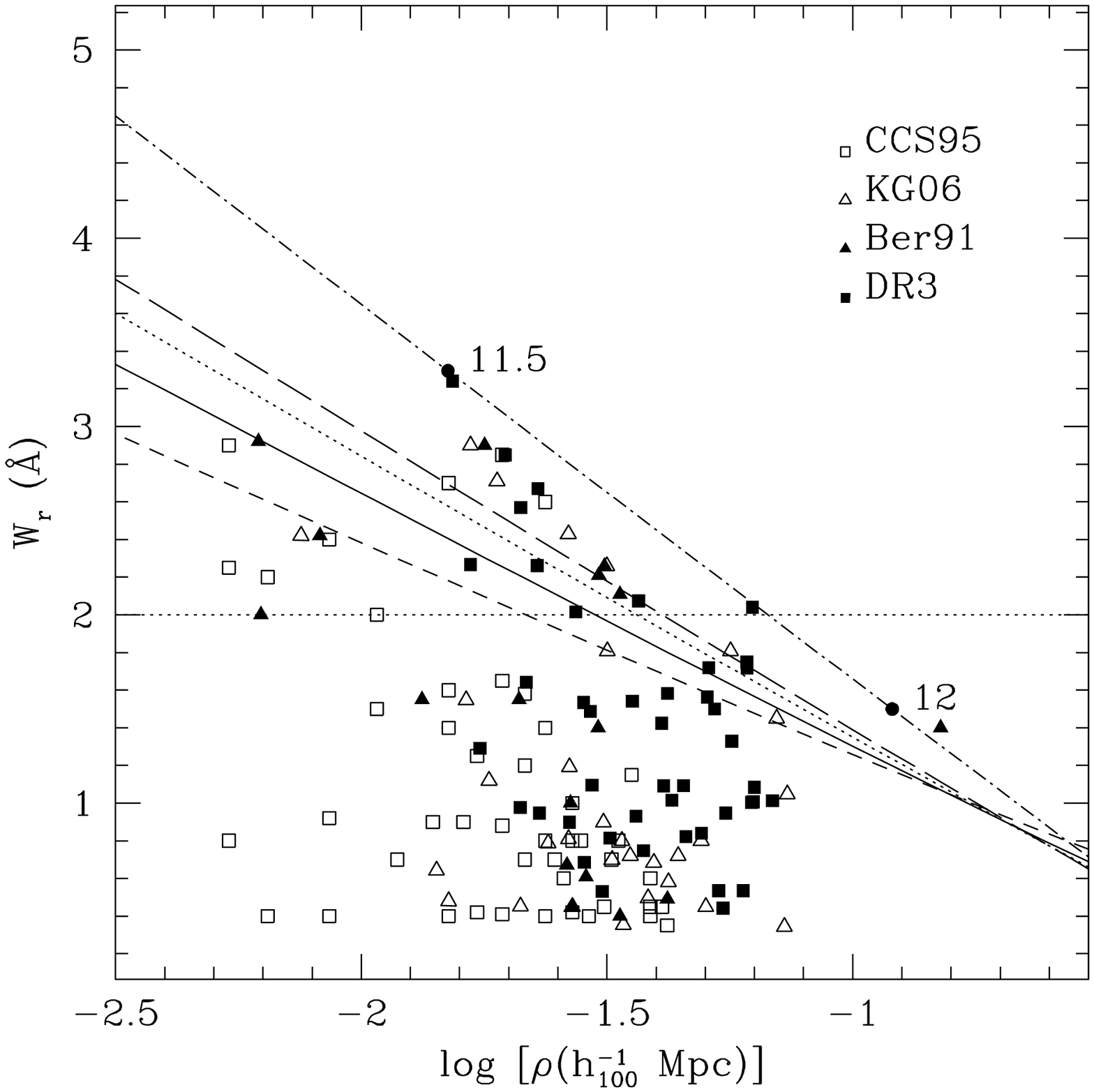}
}
\caption{{\it Left}: The black squares show the halo-mass ($\Mh$) as a
  function of equivalent width (\EW) as in
  Fig.~\ref{fig:xcorr:EW}(right). The grey squares show a different
  binning.  The various lines represent the expected \Mh--\EW\
  relationship, normalized at $\log \Mh(\msun)=12$, for various LFs
  ($\alpha=-1.0, -0.5$) and mass-to-light ratios ($q=0, -0.5, -0.75$)
  [see text and equation~(\ref{eq:predict:mass-EW})].  The results
  from our cross-correlation analysis follow the expected
  anti-correlation.  The error bars represent the 1-$\sigma$
  statistical uncertainties and the size of the equivalent width bin
  used.  {\it Right}: \EW\ as a function of impact parameter, $\rho$, for
  the \MgII\ host-galaxies.  The data points are as in
  Fig.~\ref{fig:impact-EW}.  The lines are as in the left panel and
  represent the radii of the cross-section $R_{\rm phys}$ (i.e.
  maximum impact parameter allowed).  The numbers along the dot-dash
  line indicate the logarithm of the halo mass corresponding to the \EW\
  from the left panel.  }
\label{fig:test:lum}
\end{figure*}

One can go a step further and `predict' the \Mh--\EW\ relationship
using the mass-to-light ratio.  Since the faint-end slope of the mass
function of haloes \citep[e.g.][]{MoH_02a} is steeper than that of
galaxies, the halo mass-to-light ratio, $\Mh/L$, decreases with
increasing $L$ \citep[see][and references therein]{ShankarF_06a}.
  Parametrizing $\Mh/L$ as $L^{q}$ (with $q\la0$), one
finds that the expected slope of the $\Mh$--\EW\ relationship is:
\begin{eqnarray}
\frac{\rm d\log M}{\rm d W_{\rm r}}&=& (1+q) \frac{\rm d\log L}{\rm d W_{\rm r}}\nn\\
&\simeq&(1+q)\frac{-\log_{10}(e)/W^*}{1+\alpha+2\beta}\label{eq:predict:mass-EW}
\end{eqnarray}
Fig.~\ref{fig:test:lum}(left) plots the expected relationship between
$\Mh$ and \EW\ for $q=0, -0.5$ \& $-0.75$ and $\alpha=-1.0$ \& $-0.5$,
with all curves normalized at $\log \Mh(\msun)=12$.  The data points
are as in Fig.~\ref{fig:xcorr:EW}(right) and agree well with the
expected relationships (lines).

Our conclusion from this exercise is that any plausible LF, combined
with \dNdzdW\ \citep[e.g][]{NestorD_05a} and with the observed
$R_{\rm phys}(L_K)$ relation \citep{SteidelC_95b}, together {\it
  require} a $L$--\EW\ (or \Mh--\EW) anti-correlation.

Since a $\Mh$--\EW\ anti-correlation is required from past results, we
can perform a simple consistency check as follows. We use the series
of relationships shown in Fig.~\ref{fig:test:lum}(left) and the
halo-mass function, $n(\Mh)\equiv{\mathrm d N}/{\mathrm d V}/{\mathrm d
  \log M}$, from \citet{MoH_02a} in equation~(\ref{eq:dNdz:diff}) to
predict the radius $R_{\rm phys}$ of the cross-section as a
function of equivalent width, $R_{\rm phys}(\EW)$, that is required
by the incidence distribution \dNdzdW\ of \citet{NestorD_05a}.  The
result is shown in Fig.~\ref{fig:test:lum}(right) for each of the
relationships on the left plot. The data points are as in
Fig.~\ref{fig:impact-EW}.

Thus, we have shown with two different arguments (in Sections 5.1 \&
5.2) that a positive $\Mh$--\EW\ correlation is not consistent with
the impact parameter distribution ($\rho$--\EW) of current \MgII\
samples. This conclusion is completely independent of our clustering
results presented in Section \ref{section:results}. One can turn these
arguments around and state that the anti-correlation between \EW\ and
maximum impact parameter (Fig.~\ref{fig:impact-EW}) is a natural
outcome of the \Mh--\EW\ anti-correlation in Fig.~\ref{fig:xcorr:EW}.

\section{Physical Interpretation}
\label{section:interpretation}

The \Mh--\EW\ anti-correlation has a direct and important consequence:
As described in the introduction, if the individual sub-components or
clouds of strong \MgII\ systems were virialized within the host-galaxy
halo, the velocity spread $\Delta v$ (as traced by \EW) would be a
measure of the galaxy velocity dispersion and would be correlated with
\Mh\  via $\EW/\ma=\Delta v/c$ 
 (or with the circular velocity of the galaxy, $V_c$, via
$\EW/\ma=V_c/c$).  However, as demonstrated in Sections~5.1 and 5.2,
one would not reproduce the observed relation between $R_{\rm
  phys}$ and \EW.  Thus, both our cross-correlation results and the
arguments presented in Section~5 show that the \MgII\ clouds are not
virialized in the gaseous haloes of the host-galaxies, in the sense
that the velocity spread $\Delta v$ (as traced by \EW) of the
sub-components is not proportional to halo mass.

We now turn towards a physical interpretation of our
results. In particular, can we put constraints on the competing
models for \MgII\ absorbers?

\subsection{Basic \MgII\ cloud properties}\label{section:clouds}

In a few cases the physical properties (sizes, densities etc.) of the
individual components of \MgII\ systems with $\EW>0.3$~\AA\ have been
constrained from photo-ionization modelling.  For instance,
\citet{DingJ_03a} and \citet{MasieroJ_05a} found that individual
\MgII\ velocity components (or clouds) of systems with $\EW>0.3$\,\AA\
have gas densities of $n_{\rm H}=0.001$--$0.1{\rm \,cm}^{-3}$ and,
therefore, typical sizes of $s=10$--$1000{\rm \,pc}$, 
close to the direct size measurements of \citet{RauchM_02a} using lensed QSOs.
For a warm
($T\!\sim\!10^4$\,K) cloud of size $s\!=\!100$\,pc and $n_{\rm
  H}=0.01$\,cm$^{-3}$, the sound crossing time is $\sim\!10^7$\,yr,
and the cooling time-scale is $k\,T/[n\Lambda(T)]\sim 10^6$\,yr since
the cooling function $\Lambda(T)\simeq
10^{-22}$\,erg\,cm$^3$\,s$^{-1}$ at $T=10^4$\,K
\citep{SutherlandR_93a}.  The same ionization studies indicate that
the associated \MgI\ components are denser ($n_{\rm
  H}\ga$10\,cm$^{-3}$) but very small, only $0.01$--$0.1$\,pc.  This
supports the view in which the clouds would condense out of the hot
gas.

For such warm clouds embedded in a hot ($T\sim10^6$\,K) ionized
medium, the evaporation time scale is $>\!10^7$\,yr
\citep{CowieL_77a,McKeeC_77a}.  They are stable against gravity
because their Jeans lengths are much larger than the cloud sizes, but
they would be destroyed by shocks in $s/v_{\rm shock}$$\sim1$\,Myr for
a shock velocity $v_{\rm shock}$ of 100\,\kms\ \citep{KleinR_94a}.

\subsection{HVCs and \MgII\ clouds}
 
The properties of the Galactic high-velocity clouds (HVCs) -- their
\HI\ column densities, sizes, densities etc.~-- are akin to those of
Lyman limit and \MgII\ absorbers \citep[e.g.][]{CharltonJ_00a,ZwaanM_00a}.  For
instance, \citet{RichterP_05a} studied the Galactic gaseous halo in
high-resolution spectra towards one line-of-sight (PKS 1448$-$232) and
detected low column density, high-velocity \CaII\ 
($N_{\rm CaII}\sim 10^{11\hbox{--}12}$\,\cmsq) and \NaI~D ($N_{\rm NaI}\sim 10^{11}$\,\cmsq) 
absorption at $v_{\rm LSR}\simeq-150$\,\kms\ with
Doppler widths of 4--8\,\kms. This is very similar to typical values
for \MgII\ absorbers \citep{ChurchillC_03a}. Their follow-up VLA 21-cm
studies unveiled several high-velocity \HI\ clumps at peak column
densities $\NHI\sim 10^{19}$\,\cmsq\ near the QSO line-of-sight.  The
sight-line seems to have intercepted the outer part of one of the
clumps.  Assuming a distance of 4--12\,\hkpc, the density of this
clump is $n_{\rm H}\simeq0.1$~cm$^{-3}$, similar to that inferred for
\MgII\ clouds by \citet{DingJ_03a}.  Because 8 of 13 sight-lines have
similar high-velocity \CaII\ clouds, \citet{RichterP_05a} concluded
that the covering factor of such low column density gas must be large.
Based on this they argued that if such clumps are typical for haloes
of quiescent spiral galaxies, such low column density clouds should
contribute significantly to the population of \MgII\ absorbers and
Lyman limit systems.  In a starburst environment these clouds may be
either more numerous and/or have larger column densities.

Towards the Andromeda galaxy (M31), \citet{ThilkerD_04a} reported a
population of faint \HI\ clouds detected directly with the Green Bank
Telescope (GBT).  Follow-up studies by \citet{WestmeierT_05a} using
the Westerbork telescope clearly show that the clouds have typical
sizes of $\sim 1$~kpc, \HI\ column densities of $10^{19\hbox{--}20}$\,\cmsq,
densities $n_{\rm H}\sim10^{-2}$~cm$^{-3}$, \HI\ line widths of
$\sim\!20$\,\kms\ and velocity gradients of several \kms.  The \HI\
mass of these clouds is then $\sim3\times 10^5$\,\msun.
Thus, the properties of HVCs are very close to those inferred for
\MgII\ clouds from photo-ionization studies discussed in Section
\ref{section:clouds} \citep[see also the recent analysis
of][]{FoxA_05a}.

\subsection{Finding the right model}

There are three main hypotheses that could describe \MgII\ absorbers.
The first assumes that QSO sight-lines probe galactic gaseous disks
and, therefore, the interstellar medium of the host-galaxies.  The
second assumes that each QSO sight-line probes gaseous haloes of
galaxies where the gas originates from the filaments, falls into the
galactic potential well and eventually cools onto a rotationally
supported disk. The third hypothesis places the \MgII\ absorbing gas
in the same gaseous halo, but the gas comes from outflows driven by
the large number of supernovae in star-bursts.  This last hypothesis
has long been used to model the HVCs in the context of a `galactic
fountain' scenario in a multi-phase medium
\citep[e.g.][]{ShapiroP_76a,WolfireM_95a}.

The disk model predicts a strong dependence of \EW\ on inclination of
the disk with respect to the line-of-sight.
\citet{KacprzakG_05a,KacprzakG_06a} did not find any such correlation.
Furthermore, \citet{SteidelC_02a} found that, in a small sample of
5 \MgII\ absorbers, an extension of the
host-galaxy  rotation curve  did not reproduce the absorption-line kinematics.
 A simple disk model therefore seems unlikely.

The evolution of the \MgII\ incidence probability is observed to
strongly mimic the evolution of the star formation rate density (SFRD)
with redshift, as pointed out by \citet{ProchterG_05a} and also 
\citet{MisawaT_05a}. Indeed, removing the
cosmological dependence (i.e.~$\mathrm d r/\mathrm d z$) from the
incidence probability, \dNdz, in equation~(\ref{eq:dNdz:diff}) leads to
the line-density of absorbers per co-moving Mpc, ${\mathrm
  dN}/{\mathrm dX} \equiv n(M)\,\sigma_{\rm co}$. This line-density
shows a decline at $z<1$, just like the SFRD, and both ${\mathrm d
  N}/{\mathrm d X}$ and the SFRD are approximately constant from
$z\sim3$ to $z=1$. The decline of ${\mathrm dN}/{\mathrm dX}$ can only
be explained by a decrease in the cross-section, $\sigma_{\rm co}$,
with time since the number density of haloes, $n(M)$, increases with
time for all masses.  From these arguments, \citet{ProchterG_05a}
concluded that ``the processes responsible for strong \MgII\ absorbers
are turning off at $z\sim1$''. This behaviour is natural within a
starburst scenario where \MgII\ absorbers are directly related to
supernovae-driven winds.

Regarding the in-fall hypothesis, \citet{MoH_96a} produced a model for
Lyman limit systems, \MgII\ and \CIV\ absorbers in a CDM cosmology.
This model consists of a two phase medium for the gaseous halo.  In
the context of galaxy formation in dark matter haloes
\citep[e.g.][]{WhiteS_78a,WhiteS_91a}, the initial gas in a galaxy
(either at the time of collapse or through accretion) is shocked at
around the virial radius. A halo of hot gas ($T\!\sim\!10^6$\,K) is
formed which has a temperature close to the halo virial temperature
since it is in an approximate hydrostatic equilibrium with the dark
matter. Due to cooling from thermal instabilities and inhomogeneities,
a warm phase ($T\!\sim\!10^4$\,K) will form, comprising photo-ionized
clouds confined by the pressure of the hot phase, while accreting onto
the galactic central region.

As a consequence, in this inflow picture the photo-ionized clouds will
be approximately virialized in the gaseous halo. \citet{MoH_96a}
argued that the terminal velocity of these clouds will be close to the
virial velocity for haloes with $V_c<250$\,\kms. In this scenario, the
velocity spread of the gas $\Delta v$ (as measured by \EW) would be
related to the virial velocity and therefore its halo-mass.  Thus, we
view the observed \Mh--\EW\ anti-correlation as strong evidence
against this inflow picture.  Furthermore, more massive (and larger)
galaxies with large \EW\ would be seen at large impact parameters,
contrary to the observations of \citet{LanzettaK_90a} and
\citet{SteidelC_95a}.

There is one main caveat to this conclusion. The in-fall model of
\citet{MoH_96a} might not be directly relevant to our absorbers with
$\EW\!\sim\!1$--$4$\,\AA\ which are expected to be optically thick,
i.e.~saturated. The in-fall model assumes the clouds are optically
thin, i.e.~$\EW\la0.3$\,\AA, still on the linear part of the curve of
growth. Most such absorbers are predicted to be in massive galaxies,
with a median $V_c=220$\,\kms\ [$\log \Mh(\msun)\sim {12.4}$] and most
systems in the range $V_c=150$--$300$\,\kms\ 
[$\log \Mh(\msun)\sim 11.9$--$12.8$]. 
 This mass range, predicted for \EW$\simeq0.3$\,\AA\
absorbers, agrees with our mass estimates for the lowest \EW\ systems
in our sample (0.3--1.0\,\AA). The inflow model could therefore
describe \MgII\ absorbers with $\EW\la0.3$\,\AA. We also note that the
\EW\ distribution function has different power-law indexes below and
above 0.3\,\AA\ \citep{NestorD_05a}, possibly reflecting two different
physical origins.

Another possible caveat is the recent realisation that, for low mass
galaxies, a significant fraction of the gas accretion occurs without
the initial shock heating because the cooling time is very short,
i.e.~the gas remains warm ($T\sim10^4$~K) as it moves from the
intergalactic medium to the central parts of galaxies. For this
reason, it is often referred to as the `cold mode' of gas accretion
\citep[e.g.][]{BirnboimY_03a,KeresD_05a}.  In this context, the clouds
will not be virialized, but the terminal velocity should still be
related to the halo-mass.

\citet{MallerA_04a} extended the arguments of \citet{MoH_96a} to
include multi-phase cooling since the hot gas in galactic haloes is
thermally unstable (see also discussion in Section
\ref{section:clouds}). This multi-phase cooling scenario allows the
survival of a hot gas core and naturally gives rise to the formation
of $\sim\!10^4$~K clouds, each of mass $\sim10^6$~\msun, which are
pressure-confined by the hot halo.  The main conclusion of this model
relevant to our work is that this multi-phase treatment naturally
explains the properties of \MgII\ clouds and HVCs: the unstable hot
gas produces warm clouds of mass $\sim\!10^6$\,\msun\ which are
stabilized by the pressure of the surrounding hot gas. In this model,
the maximum distance or impact parameter for warm clouds (HVCs) is set
by the the `cooling radius', which is $\sim\!100$\,kpc for a Milky Way
size halo.  This model differs from the one of \citet{SternbergA_02a}
in that the self-gravity plays a negligible role and no dark matter is
required.

\citet{SternbergA_02a} modelled the HVCs as multi-phase (warm-ionized
and cold-neutral) pressure-confined cloud, each belonging to  a `mini-halo' made of dark matter.
In this model, the dark matter in the mini-haloes provide the extra pressure
in the core of the clouds required to allow for the formation of a
cold medium.  
The warm phase of the HVCs is confined by the ambient pressure
provided by the hot ($T\!\sim\!10^6$~K) ionized gas of the halo. 
This hot gas is in hydro-static equilibrium with the
dark-matter of the host galaxy and determines the pressure profile of
a galaxy. Given the observed velocity widths (10--20\,\kms),
\citet{SternbergA_02a} determined that the pressure of this hot-gas
must be $P/k>50$\,cm$^{-3}$\,K to keep the warm phase of the clouds
bound. In the Milky Way, the
pressure of the hot halo is above $50$\,cm$^{-3}$\,K out to
150--200\,\hkpc, i.e.~almost to the virial radius. For galaxies with
smaller halo-mass (or smaller virial temperature), the central
pressure will be reduced and the radius at which $P/k$ reaches
50\,cm$^{-3}$\,K will be smaller than the virial radius, at which
point the cloud will evaporate.

\begin{figure}
\centerline{\includegraphics[width=90mm]{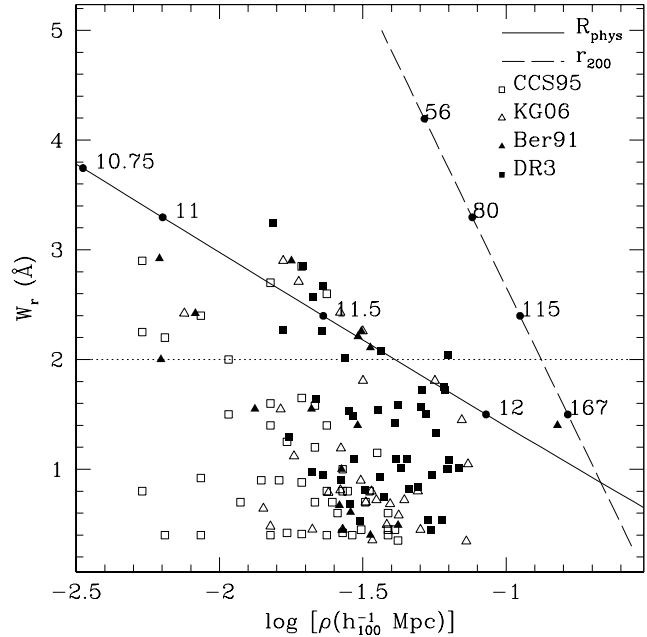}}
\caption{Equivalent width, \EW, as a function of impact parameter,
  $\rho$, as in Fig.~\ref{fig:test:lum}(right).  The data points are as
  in Figs.~\ref{fig:impact-EW} \&~\ref{fig:test:lum}(right).  The
  solid line shows one of the model curves from
  Fig.~\ref{fig:test:lum}, the long dashed line.  The numbers along
  the solid line indicate the logarithm of the halo mass in units of
  \msun) corresponding to the equivalent width.  The dashed line shows
  the virial radius of such haloes ($r_{200}$), with the corresponding
  virial velocity marked. The virial radius is much greater than the
  cross-section radius, $R_{\rm phys}$, for absorbers with
  $\EW\!>\!1.5$\,\AA. Therefore, strong \MgII\ absorbers seem not to
  fill the haloes of their host-galaxies.
  }
\label{fig:r200}
\end{figure}

The fact that no absorbers with large \EW\ are found to have
host-galaxies at large impact-parameters (i.e.~no points to the right
of the solid line in Fig.~\ref{fig:impact-EW}) clearly tells us
something about the characteristic scale where clouds either form or
evaporate, and not the size of the halo of the host-galaxy.  Indeed,
for the halo-masses found in Fig.~\ref{fig:test:lum}(left) one can
compare the corresponding virial radii to the $\rho$--\EW\ distribution
in Fig.~\ref{fig:test:lum}(right), as we show in
Figure~\ref{fig:r200}. The dashed line represents the $z=0.5$ virial
radius, $r_{200}$, for the model with $\alpha=-0.75$ and $q=0$ in
Fig.~\ref{fig:test:lum}(left). The logarithm of the halo-masses (in \msun)
 are marked along the solid line. The halo circular velocity, $V_c$, is marked along the
dashed line in \kms. This figure clearly shows that $r_{200}$ is much
larger than the cross-section radius ($R_{\rm phys}$, solid line)
for absorbers with $\EW\!>\!1.5$\,\AA\ (or halo-masses
$M_h\!<10^{11.5}$\,\msun). In other words, \MgII\ absorbers do not fill the 
halo of the host-galaxy.
The radius of the cross-section must then  be due to
some other physical mechanism. It could be 
either the pressure threshold discussed by \citet{SternbergA_02a} and/or
to the cooling radius of \citet{MallerA_04a}.

\subsection{In the context of super-winds}
\label{section:context}

Local supernovae-driven winds are known to expel ionized gas into
their haloes, as traced by \Ha, X-rays and certain emission lines
\citep[e.g.][]{LehnertM_94a,HeckmanT_95a,ArmusL_95a,LehnertM_96a,WangJ_97a,DahlemM_97a}.
However, the presence of \NaI\ D absorption also indicates that the
winds carry large amounts of dust and cold gas, $M\ga10^8\msun$, over
the life-time of the star-burst
\citep[e.g.][]{LehnertM_96a,MartinC_99a,HeckmanT_00a,RupkeD_05a,MartinC_06a}.  The dust can
sometimes be seen in direct imaging \citep{HowkJ_00a} out to a few
kpc.  Furthermore, the amount of dust is correlated with the amount of
cold gas traced by \NaI~D absorption \citep{HeckmanT_00a,MartinC_06a}.  This shows
that cold material is entrained and expelled outside the disk by the
ionized gas flow.  In this context, the findings of
\citet{KondoS_06a}, \citet{YorkD_06a} and \citet*{WildV_06a} are very
interesting. For the first time, \citeauthor{KondoS_06a} detected
strong \NaI~D$\lambda\lambda$5891/5897 doublet absorption in two strong
\MgII\ absorbers (also DLAs) at $\zabs=1.062$ \& $1.18$ towards the
gravitationally lensed QSO APM 08279$+$5255. \citet{YorkD_06a}
constrained the dust-reddening from the slope of stacked QSO spectra
with foreground \MgII\ absorbers selected similarly to ours, but at
$1\!<\!\zabs\!<\!1.8$. Finally, \citet{WildV_05a} and \citet{WildV_06a}
found that the sub-sample of strong \MgII\ absorbers showing
significant \CaII\ absorption in the low-resolution SDSS QSO spectra
are even more dust-reddened. Both \citeauthor{YorkD_06a} and
\citeauthor{WildV_06a} found that the reddening is correlated with
equivalent width, a relationship very similar to that of
\citet{HeckmanT_00a}.

The metallicity of our \MgII\ systems is unknown but
\citeauthor{YorkD_06a} used indirect arguments to estimate average
metallicities for a DR1 sample of \MgII\ absorbers. They found that
the metallicity decreases with increasing \MgII\ equivalent
width\footnote{Using a very different method to estimate \NHI,
  \citet{NestorD_03a} found a correlation between \EW\ and the
  metallicity of DLAs selected with the \MgII\ criteria of
  \citet{RaoS_00a} and \citet{RaoS_06a}. As noted by
  \citeauthor{NestorD_03a}, this relationship can be understood if
  higher \EW\ absorbers correspond to regions with more intense bursts
  of star formation.}. This would lead to a normal mass--metallicity
relation when combined with our \Mh--\EW\ anti-correlation.

Given these local results and that \MgII\ absorbers predominantly
trace clouds of `cold' material, are the \MgII\ clouds related to the
entrained material in outflows~\footnote{After completion of this work,
\citet{MartinC_06a} argued that the
relics of cool outflows of Ultraluminous Infrared Galaxies (ULIRGs) and Lumious Infrared galaxies (LIRGs),
as traced by NaI~D, will create a significant redshift-path density for \MgII\
if most $L \!>\! 0.1\!L^*$ galaxies pass through a luminous starburst phase.}
? If so, then \EW\ should be a measure
of the (past) star-formation rate and mass ejection history for the following
reasons.  Firstly, \EW\ is a measure of the number of components (or
clouds), as shown by \citet{ChurchillC_03a}.  Secondly, our results
show that absorbers with $\EW\ga 2$\,\AA\ have smaller potential wells
so the host-galaxy's supernovae-driven winds are more likely to eject
gaseous material out of the disks. 
Thus, in the context of galactic
outflows, the velocity spread (and therefore \EW), which is just a
measure of the number of components, is   related to the (past)
star-formation rate and to the mass ejection rate.

Entrained clouds would have to survive from the disk to where they are
seen at $R\sim50$\,kpc. The travel time to 50\,kpc is about
100\,Myr for wind speeds of 500\,\kms.  From the discussion in Section
\ref{section:clouds}, the formation and destruction time-scales are
1--2 orders of magnitude smaller than the travel time or halo
dynamical time ($\simeq$~0.5--1\,Gyr).  It seems very unlikely that
such clouds could survive several times 100\,Myr given these physical
properties unless they are confined by the pressure of the hot ionized
medium. The same conclusion was reached by \citet{SternbergA_02a} and
\citet{MallerA_04a} and is discussed above.

Recently, \citet{KacprzakG_05a,KacprzakG_06a} reported that the
host-galaxy morphological asymmetry and \EW\ are correlated. While
these authors favour the merger interpretation for their results, they
also discuss the possibility of star-burst driven winds. Again,
because of the smaller potential wells of higher \EW\ absorbers,
outflows driven by spatially asymmetric star-bursts can create
asymmetries and produce the correlation seen by
\citet{KacprzakG_05a,KacprzakG_06a} without invoking mergers. In fact,
one sees that most \MgII\ host-galaxies in their sample are relatively
isolated and do not show galaxy pairs or strongly disturbed
morphologies.
 
\subsection{Further evidences for the super-wind scenario}

Direct evidence of outflows exists in only a few \MgII\ absorbers.
\citet{BondN_01a} presented high-resolution spectra of four strong
($\EW\!>\!1.8$\,\AA) \MgII\ absorbers at $1\!<\!z\!<\!2$ and showed
that the profiles ``display a common kinematic structure, having a
sharp drop in optical depth near the centre of the profile and strong,
black-bottomed absorption on either side''. They interpret these
features as a signature of super-winds arising in actively
star-forming galaxies. In the similar case of the $z=0.74$ \MgII\ host-galaxy
towards QSO Q1331+17 \citep{EllisonS_03a}, the absorber's absolute
luminosity is quite low, $L\sim 0.3L^*$, and the authors suggest that
``the double-peaked absorption profiles, their striking symmetries,
...  and large \MgII\ equivalent width support the interpretation of
superbubbles''.

\citet{NormanC_96a} searched for \MgII\ absorption in QSO spectra that
lie near the starburst galaxy NGC 520, a well-known local starburst
with super-winds. NGC 520 has an IR-luminosity comparable to that of
M82, shows a disturbed morphology and strong filamentary \Ha\ emission
along the minor axis. In \citeauthor{NormanC_96a}, the \MgII, \MgI,
and \FeII\ absorption associated with its gaseous halo are clearly
detected along two lines-of-sight. The QSOs Q0121+0338 and Q0121+0327
intersect the halo of NGC 520 at $\sim\!55$\,\hkpc\ and
$\sim\!25$\,\hkpc, respectively. In both sight-lines, the \MgII\
equivalent widths are greater than $0.3$\,\AA: $\EW=0.33$\,\AA\
towards Q0121+0338 and $\EW=1.7$\,\AA\ towards Q0121+0327.  The latter
absorber with the largest \EW\ has the smallest impact parameter, as
expected from the larger sample in Fig.~\ref{fig:impact-EW}.


\section{Summary}\label{section:summary}
 
From the SDSS/DR3 we selected \nabstot\ \MgII\ absorbers with
$\EW\!\geq\!0.3$\,\AA\ in the redshift range
$0.4\!\leq\!\zabs\!\leq\!0.8$ and $\ngalsim$ LRGs in the absorber
fields selected with photometric redshifts. From the ratio of the
\MgII--LRG cross-correlation to the LRG--LRG auto-correlation, we
constrained the halo-mass of \MgII\ absorbers in a statistical manner,
i.e.~without directly identifying them.  To summarise our main
results, we have shown that:
\begin{enumerate}
\item the ratio of the cross- and auto-correlation, $w_{\rm ag}/w_{\rm
    gg}$, is $a=\Arel$;
\item the corresponding bias-ratio between strong \MgII\ absorbers and
  LRGs is $b_{\rm \MgII}/b_{\rm LRG}=\Arelcorr$;
\item this bias-ratio implies that the absorber host-galaxies have a
  halo-mass of $\langle \log \Mh (\msun)\rangle =\;$\LRGmassrangesys.
  The 1-$\sigma$ uncertainty includes the statistical errors in $a$
  and possible systematic errors in the LRG halo-mass;
\item the \MgII\ equivalent width, \EW, is significantly
  anti-correlated with the absorber halo-mass, \Mh.
\end{enumerate}
The main consequence of point (iv) is that the \MgII\ clouds are not
virialized in the gaseous haloes of the host-galaxy, in the sense that
$\Delta v$, the velocity spread of the individual clouds (ranging from 50 to $\sim~400$~\kms), does not
positively correlate with halo mass. In other words, had the clouds
been in virial equilibrium with the halo dark matter distribution,
$\Delta v$ (as measured by \EW) would correlate with the mass of the
halo. However, we find the opposite: lower-mass haloes harbour the
host-galaxies producing the largest number of individual \MgII\ clouds
spread over the largest velocities.

Since the \Mh--\EW\ anti-correlation may initially seem surprising, we
performed several consistency tests on the method itself and tested
the \Mh--\EW\ anti-correlation against numerous past results in the
literature. For the galaxy clustering method we have shown that:
\begin{itemize}
\item The cross-correlation is solely due to the \MgII\ absorbers;
\item The cross-correlation amplitude decreases when the width of the
  LRG redshift distribution is increased, as expected;
\item The $w_{\rm ag}/w_{\rm gg}$ ratio is independent of the LRG
  redshift distribution, as expected if one uses the same galaxies to
  calculate both $w_{\rm ag}$ and $w_{\rm gg}$. Therefore, the
  relative amplitude is free of possible systematic errors caused by
  contaminants (stars or interloping galaxies).
\end{itemize}

In checking past results in the literature for consistency with the
observed $\Mh$--\EW\ anti-correlation we found that:

\begin{itemize}
\item When one combines the observed luminosity- and \EW-dependence
  of the cross-section, one finds that luminosity (or mass) and
  equivalent width must be anti-correlated;
\item When one combines the observed luminosity-dependence of the
  cross-section with the observed incidence probability of strong
  \MgII\ absorbers, \dNdzdW, one finds that luminosity (or mass) and
  equivalent width must be anti-correlated for all plausible
  luminosity functions with faint-end power-law indexes $\alpha >
  -1.4$.
\end{itemize}
In other words, a positive correlation between \EW\ and halo-mass (or
virial velocity) -- expected if the clouds were virialized -- is
inconsistent with past results. This is independent of the \Mh--\EW\
anti-correlation from the clustering analysis. When one combines the
observed \Mh--\EW\ relationship with the observed \dNdzdW\ one
predicts a cross-section radius which agrees very well with the
distribution of absorber impact parameters observed in previous and
current samples.

We interpret our results as a strong indication that a large
proportion of strong \MgII\ absorbers, particularly those with $\EW\ga
2$\,\AA, arise in galactic outflows.
A conclusion similar was reached by \citet{SongailaA_06a} for $z\sim2$ \CIV\ absorbers.
Supernovae-driven winds drive hot
gas out of the galaxy plane, forming cavities if numerous supernova
explosions occur in a small region on a similar time-scale. The hot
gas breaks out of the disk, forming a hot gaseous corona (traced by
high-ions such as \CIV\ and \OVI) and may entrain cold, dusty
material.  Part of this gas cools, forming pressure-confined clouds
traced by \MgII\ which may be the analogs of the HVCs.

This interpretation allows us to make several predictions.  Firstly,
it provides a natural explanation as to why absorbers with
$\EW>2$\,\AA\ are not observed at large impact parameters. These
systems should be physically closer to the host galaxy and also closer
in time to more recent and more active star formation. Thus, they
ought to have star formation rates typical of star-bursts. The
resonance line O[{\sc ii}]$\lambda$3727 may also be detectable in
stacked SDSS spectra since the host-galaxies of $\EW\ga 2$\,\AA\
absorbers will be very close to the line-of-sight. The detectability
of this emission line should be a strong function of \EW\ but will
also obviously depend on the host-galaxy metallicity.  
Secondly, the
\NaI~D doublet ought to be detectable in the same stacked spectra.
Thirdly, the colours of the stacked light distribution should be a
strong function of \EW. The preliminary results of S.~Zibetti (private
communication) confirm this last prediction.

\section*{Acknowledgments}

We would like to thank the anonymous referee for his/her comments which
led to an improved manuscript.  We thank J.~Bergeron who pointed out
the wind scenario; C.~W.~Churchill \& M.~Zwaan for discussions on
\MgII\ absorbers; M.~D.~Lehnert on the physics of winds.  H.-W.~Chen
\& M.~Zwaan provided us with detailed comments on an early manuscript,
which led to several improvements in the presentation of the results.
G.~Kacprzak kindly provided us with the impact parameters for his
sample prior to publication.  M.~T.~Murphy thanks PPARC for an
Advanced Fellowship.  I.~Csabai acknowledges the following grants:
OTKA-T047244, MSRC-2005-038 and MRTN-CT-2004-503929. Funding for the
Sloan Digital Sky Survey (SDSS) has been provided by the Alfred
P.~Sloan Foundation, the Participating Institutions, the National
Aeronautics and Space Administration, the National Science Foundation,
the U.S.~Department of Energy, the Japanese Monbukagakusho, and the
Max Planck Society.


\bsp_small

\label{lastpage}

\end{document}